\documentclass{article}

\usepackage{microtype}
\usepackage{graphicx}
\usepackage{subfig}
\usepackage{booktabs} 
\usepackage{amsmath}
\usepackage{dsfont}
\usepackage{cuted}

\usepackage{hyperref}

\usepackage[accepted]{icml2021}

\icmltitlerunning{Prediction-Centric Learning of Independent Cascade Dynamics from Partial Observations}
\begin{document}

\twocolumn[
\icmltitle{Prediction-Centric Learning of Independent Cascade Dynamics\\ from Partial Observations}



\icmlsetsymbol{equal}{*}

\begin{icmlauthorlist}
\icmlauthor{Mateusz Wilinski}{lanl}
\icmlauthor{Andrey Y. Lokhov}{lanl}
\end{icmlauthorlist}

\icmlaffiliation{lanl}{Theoretical Division, Los Alamos National Laboratory, Los Alamos, USA}
\icmlcorrespondingauthor{Mateusz Wilinski}{mateusz@lanl.gov}
\icmlcorrespondingauthor{Andrey Y. Lokhov}{lokhov@lanl.gov}

\icmlkeywords{dynamic message passing, learning, inference, networks, independent cascade model}

\vskip 0.3in
]



\printAffiliationsAndNotice{}  

\begin{abstract}
Spreading processes play an increasingly important role in modeling for diffusion networks, information propagation, marketing and opinion setting. We address the problem of learning of a spreading model such that the predictions generated from this model are accurate and could be subsequently used for the optimization, and control of diffusion dynamics. We focus on a challenging setting where full observations of the dynamics are not available, and standard approaches such as maximum likelihood quickly become intractable for large network instances. We introduce a computationally efficient algorithm, based on a scalable dynamic message-passing approach, which is able to learn parameters of the effective spreading model given only limited information on the activation times of nodes in the network. The popular Independent Cascade model is used to illustrate our approach. We show that tractable inference from the learned model generates a better prediction of marginal probabilities compared to the original model. We develop a systematic procedure for learning a mixture of models which further improves the prediction quality.
\end{abstract}

\section{Introduction}

Spreading processes on networks have seen an increasing interest in the domains related to modeling of infectious diseases, regulatory networks, marketing and opinion dynamics. In a realistic setting, the model and parameters, as well as propagation paths are unknown, and one is left at most with several observed diffusion traces in a form of nodes' activation times. Moreover, even this information is not fully available since the time-stamps of node activations may not be available from each node in the network. This motivates the development of efficient algorithms that can infer the effective spreading models from incomplete information.

In the recent years, the problem of diffusion model recovery from full observations has been extensively addressed through a series of works focusing on heuristic and exact algorithms, largely based on maximum likelihood type approaches \cite{netrapalli2012learning, gomez2012inferring, gripon2013reconstructing, abrahao2013trace, pouget2015inferring, braunstein2019network}. These methods allow one to recover the parameters of the model, as well as the structure of the diffusion graph. Comparably, much less has been done on the side of parameter recovery in the case of partial observations. This is primarily due to the hardness of the estimation in this case, where maximum likelihood marginalized over the hidden observations quickly becomes computationally prohibitive with the number of hidden nodes \cite{lokhov2016reconstructing}. The work \cite{amin2014learning} primarily considered the case of tree graphs, and studied the network reconstruction problem for the case where only the initial and the final states of the dynamics are observed. This setting was further developed in \cite{supeesun2017learning} and very recently in \cite{han2020statistical} for more general graphs. In a different setting, \cite{sefer2015convex} addressed the structure learning problem in the case of missing information in time through relaxation optimization techniques under the assumption that the full probabilistic trace for each vertex is available.
The case of noise in time data was analysed in \cite{hoffmann2019learning} and \cite{trouleau2019learning}, where the latter assumed a Hawkes process driving the dynamics.
An algorithm for learning model parameters based on estimating marginal probabilities through a dynamic message-passing approach has been suggested in \cite{lokhov2016reconstructing}. Although this method provides a low estimation error and is computationally efficient compared to the marginalized likelihood, it still suffers from a computational complexity that scales quadratically in the system size for each step of the gradient descent, which limits the applicability of the method to small systems with an order of dozens of nodes only.

Much of the work above focuses on an accurate estimation of model parameters. However, in many applications the main reason for learning a model from data consists in its subsequent use for inference, optimization, or control tasks \cite{kempe2003maximizing, nowzari2016analysis}. In this setting, sometimes referred to as ``learning for inference'', or ``inferning'' \cite{heinemann2014inferning, bresler2020learning, boix2021chow}\footnote{See also the ICML 2013 workshop on ``Inferning: Interactions between Inference and Learning'': \url{http://inferning.cs.umass.edu/2012}.}, the main interest of the learning procedure is not focused on an accurate estimation of the true model parameters, but rather on the construction of an effective model such that the predictions generated from this model are accurate. The prediction task that is natural for the spreading dynamics is that of prediction of marginal probabilities of activation for each individual node that collectively quantify the expected size of the diffusion spread \cite{kempe2003maximizing}. One way to address this problem was proposed in a line of works that address the problem of direct estimation of influence functions from samples~\cite{du2014influence, NIPS2015_5989, he2016learning, makar2018learning}. However, one could be interested in constructing an explicitly parametrized influence function with an associated inference technique that can be used for other tasks beyond prediction of spread. Similarly to how belief propagation is used for approximate inference in graphical models \cite{wainwright2008graphical}, the task of an efficient approximate estimation of marginal probabilities is solved in the spreading context by dynamic message-passing (DMP) equations that avoid the computational overhead of repetitive Monte-Carlo sampling \cite{lokhov2019scalable}. Moreover, explicit equations that compute marginals has proven useful in inference tasks such as localization of the origin of the spreading process \cite{lokhov2014inferring}, or optimization tasks such as control of spreading through resource allocation \cite{lokhov2017optimal}. Therefore, it is natural to consider a learning procedure that would construct an effective model for which approximate inference techniques would yield an accurate prediction of the marginals.

In this paper, we make two contributions to the problems above. First, we revisit the objective function introduced in \cite{lokhov2016reconstructing}, and develop an exact method that computes its gradient in a time linear with respect to the dimension of the problem, thus overcoming the computational limitation of the DMPrec algorithm presented in that work. This enables application of model learning from partial observations to large realistic size networks. We also provide an interpretation of this objective function as a Kullback–Leibler divergence between empirical and model marginal distributions. This leads to the second major contribution of this work: considering dynamic message-passing as an approximate inference method, we show that inference from the effective model learned with our method generates a better prediction of marginal probabilities than inference from the original model. This observation is in agreement with previous work in learning of graphical models \cite{wainwright2006estimating, heinemann2014inferning} that showed that an inconsistent parameter estimation that takes approximate inference (such as belief propagation) into account is provably beneficial for the prediction task. We capitalize on this observation and develop a novel procedure for learning a mixture of spreading models on several replicas of the graph which further improves prediction quality of marginal probabilities. Although our approach is general and can be adapted to a large class of stochastic spreading models, we illustrate our approach on the popular Independent Cascade (IC) model \cite{goldenberg2001talk, kempe2003maximizing}.

\section{Model}

We define the diffusion network as $G = (V, E)$, where $V$ is the set of nodes and $E$ is the set of edges. In the IC model, each node can be in either of two states: active or inactive. Node $i$ activated at time $t$ has a single chance to activate its inactive neighbor $j$ at time $t+1$ with probability $\alpha_{ij}$ associated with edge $(i, j) \in E$. A cascade starts with a set of initially active nodes and continues until the dynamics dies out or at most for a predefined number of steps $T$. Since each node can be activated no more than once, every cascade $c$ can be fully described by a set of activation times $\{ \tau^c_i \}_{i \in V}$. Additionally, if node $i$ is not activated during the dynamics, we assign $T$ as its activation time by definition. We assume that we observe $M$ independent cascades with potentially different initial seeds. In general, there can be one or several seeds at initial time, or even stochastic initial conditions, where node $i$ is independently activated with probability $p^c_i(0)$ at initial time in each cascade $c$.

In practice, full observation of cascades is rarely available, and hence we split the activation times into observed and unobserved parts.
Our goal is to learn an effective dynamic model from the set of observed activation times. We will be interested in two tasks: (i) learning model parameters, i.e. estimating $\{ \alpha_{ij} \}_{(i,j) \in E}$ so that they are close to the parameters of the original model used to generate data, denoted as $\{ \alpha^*_{ij} \}_{(i,j) \in E}$; and (ii) prediction of marginal probabilities $\{ p_i(t) \}_{i \in V}$, where $p^c_i(t)$ is the probability that node $i$ is active at time $t$, so that they match the marginal probabilities of the true model denoted as $\{ p^{c*}_i(t) \}_{i \in V}$. 

\section{Methods}

In this section, we first discuss the dynamic message-passing approximate inference algorithm for estimating marginal probabilities for the IC model, and then introduce our algorithm for learning model parameters from data.

\subsection{Dynamic Message Passing Equations}

Computation of marginal probabilities of activation and of the influence function -- sum of marginals at final time $T$, which quantifies the expected spread -- is \#P-hard in the IC model  \cite{chen2010scalable, shapiro2012finding}. In practice, one could estimate the marginal probabilities using Monte-Carlo sampling, however the required number of effective runs typically scales at least linearly with the size of the system \cite{kempe2003maximizing,chen2009efficient,du2013scalable,cohen2014sketch,lucier2015influence,nguyen2017outward}. Here, we present dynamic message-passing (DMP) equations, a low complexity approximate inference algorithm that does not rely on sampling, and allows for an efficient estimation of marginal probabilities that is exact on tree graphs and asymptotically exact on random graphs.

DMP equations approximate $\{ p^c_i(t) \}_{i \in V}$ -- the set of marginal probabilities that each node $i \in V$ is active at time $t$ under a given initial condition in the cascade $c$. The computation is performed with additional quantities called messages $p^c_{j \rightarrow i}(t)$, which have a meaning of probabilities of node $j$ being active at time $t$ on an auxiliary graph where $i$ has been removed. These probabilities can be calculated by solving a following system of equations:
\begin{equation}
    p^c_i(t) = 1 - \big( 1 - \bar{p}^c_i \big) \prod_{k \in \partial i} \Big(1 - \alpha_{ki} \cdot p^c_{k \rightarrow i}(t - 1) \Big),
    \label{eq:marginal}
\end{equation}
\begin{equation}
    p^c_{j \rightarrow i}(t) = 1 - \big( 1 - \bar{p}^c_j \big) \prod_{k \in \partial j \setminus i} \Big(1 - \alpha_{kj} \cdot p^c_{k \rightarrow j}(t - 1) \Big),
    \label{eq:message}
\end{equation}
where $\bar{p}^c_i = p^c_i(0)$ is the initial probability of being active for node $i$ under initial conditions of cascade $c$.
Properties of these equations have been extensively studied in \cite{lokhov2019scalable}, and are related to the DMP equations for the popular SIR model in the limit of deterministic recovery \cite{karrer2010message, lokhov2015dynamic}, while the large-time form of these equations have been previously derived in \cite{abbe2017nonbacktracking, burkholz2019cascade}. Notice that unlike belief propagation, DMP equations do not need to be iterated until convergence and estimate marginals on any graph. Although exactness of DMP predictions are guaranteed on tree graphs only, they often provide accurate approximations on sparse but loopy graphs. Moreover, for general graphs, the solution of above equations provides an upper bound for marginal probabilities \cite{lokhov2019scalable}. Note that for given set of parameters, the solution depends only on the initial state $\bar{p}^c_i$.

\subsection{Constrained Optimization Based Learning Framework}

Next, we explain how DMP equations combined with a Lagrangian formulation of the learning problem can be used to estimate model parameters $\{ \alpha^{*}_{ij} \}_{(i,j) \in E}$ even in the case of partial observations. Our approach is based on the maximization of the following objective function:
\begin{equation}
        \mathcal{O} = \sum_{c \in C} \sum_{i \in O} \log \mu^c_i(\tau^c_i),
    \label{eq:objective_naive}
\end{equation}
where $C$ is the set of cascades, and $\mu^c_i(\tau^c_i)$ are marginal probabilities that node $i$ in a cascade $c$ is activated precisely at time $\tau^c_i$. Under the conventions introduced above, these quantities can be approximately expressed through the marginals estimated with the DMP equations as $\mu^c_i(\tau^c_i) = p^c_i(\tau^c_i) \cdot \mathds{1}_{(\tau^c_i < T)} - p^c_i(\tau^c_i - 1) \cdot \mathds{1}_{(\tau^c_i > 0)} + \mathds{1}_{(\tau^c_i = T)}$.
All cascades can be divided into classes according to their initial conditions. Let us denote the set of such classes as $S$. Notice that for given parameters, the number of different sets of messages and marginals is of the same size as $S$, because they only differ by initial conditions. As a result, we can rewrite the objective as follows:
\begin{equation}
        \mathcal{O} = \sum_{s \in S} \sum_{i \in O} \sum_{\tau_i^s} m^{\tau^s_i} \log \mu^s_i(\tau^s_i),
        \label{eq:objective}
\end{equation}
where $m^{\tau^s_i}$ is the number of cascades in class $s$ for which node $i$ was activated at time $\tau^s_i$, computed from available samples. The objective \eqref{eq:objective} is related to the Kullback-Leibler (KL) divergence between empirical and model marginal distributions (computed with DMP) for the observed nodes. The key premise of this objective function that enables reconstruction of model parameters even in the case of partial observations consists in the fact that marginal probabilities on observed nodes depend on all model parameters, including on those that are adjacent to hidden nodes. This objective has been originally proposed in \cite{lokhov2016reconstructing} with a different interpretation, with its form inspired by a mean-field like approximation to the full likelihood function. This work also included a proof of the asymptotic consistency of DMP-based recovery of model parameters using the objective \eqref{eq:objective} on tree graphs. The DMPrec algorithm proposed in \cite{lokhov2016reconstructing} consists in estimating the gradient of the objective via a direct derivative of the DMP equations. This leads to a single optimisation step complexity $O(|E|^2 T M)$, where $M$ is the number of cascades. Here, we propose a more efficient algorithm that computes the exact gradient of the objective function in time $O(|E| T M)$ at each step of the optimization, which will allow to apply it to large-sized networks.

We express the model estimation problem as a standard Lagrangian formulation for the constrained optimization problem, with the objective function \eqref{eq:objective} and the constraints expressing the fact that the marginals are estimated using DMP equations:
\begin{equation}
    \mathcal{L} = \underbrace{\mathcal{O}}_{objective} + \underbrace{\mathcal{C}}_{constrains}.
\end{equation}

The constraints are given by DMP equations (\ref{eq:marginal})-(\ref{eq:message}) re-weighted by Lagrange multipliers, where $\lambda^s_i(t)$ and $\lambda^s_{i \rightarrow j}(t)$ are Lagrange multipliers corresponding respectively to equations (\ref{eq:marginal}) and (\ref{eq:message}).

The resulting expressions for all the quantities can be found by differentiating the Lagrangian. Differentiation over marginals yields expressions for $\lambda^s_i(t)$ multipliers for all sources $s$ and times $t$. Similarly, the $\lambda^s_{i \rightarrow j}(t)$ multipliers can be obtained from the derivatives over messages, which give rise to message-passing equations in the dual space of Lagrange multipliers. Finally, derivatives over parameters $\alpha_{ij}$ take a simple form:
\begin{equation}
    \frac{\partial \mathcal{L}}{\partial \alpha_{ij}} = \frac{-1}{\alpha_{ij}} \sum_{s \in S} \sum_{t=0}^{T-1} \left(\lambda^s_{i \rightarrow j}(t) \, p^s_{i \rightarrow j}(t) + \lambda^s_{j \rightarrow i}(t) \, p^s_{j \rightarrow i}(t)\right),
    \label{eq:alpha}
\end{equation}
for non-zero parameters $\alpha_{ij}$. Now, we can use gradient components (\ref{eq:alpha}) to update parameters $\alpha_{ij}$:
\begin{equation}
    \alpha_{ij} \longleftarrow \alpha_{ij} + \varepsilon \cdot \frac{\partial \mathcal{L}}{\partial \alpha_{ij}},
    \label{eq:step}
\end{equation}
where $\varepsilon$ is a learning rate. The resulting algorithm can be expressed in a form of forward-backward iterations: (i) Start with initial parameter guess $\alpha_{ij}$; (ii) Calculate marginals and messages using forward DMP equations \eqref{eq:marginal}-\eqref{eq:message}; (ii) Use estimated marginals and messages to compute Lagrange multipliers through the ``backward'' equations; (iv) Use equations (\ref{eq:alpha}) and (\ref{eq:step}) to update parameters $\alpha_{ij}$; (v) Go back to step 2, and repeat the process with updated $\alpha_{ij}$, until the global convergence of the algorithm. The complete form of the algorithm along with the full derivation is presented in the Appendix, section \ref{appendix:1}, where we also discuss the selection of $\varepsilon$ and its relation to the convergence of our learning procedure. We refer to the above algorithm as to the Scalable Learning of Independent Cascade Effective Representation (SLICER).

Calculating marginals using DMP equations requires complexity $O(|E| T)$ and needs to be done for each different initial point of all available cascades, resulting in the worst-case computational complexity $O(|E| T M)$. In the case of single-node initial conditions, this complexity is  upper-bounded by $O(|E| T N)$, and for stochastic initial conditions is only $O(|E| T)$. Lagrange multipliers require the same computational complexity, since backward equations are equivalent to dynamic message-passing equations in the dual space. Finally, derivatives with respect to $\{ \alpha_{ij} \}_{(i,j) \in E}$, as seen from equation (\ref{eq:alpha}), are computed with the same computational complexity. As a result, overall SLICER has a complexity $\min \left[ O(|E| T M), O(|E| T \vert S \vert) \right]$. In the next section, we test the SLICER algorithm on synthetic data.

\section{Estimation of Model Parameters}

In this section, we focus on the problem of parameter estimation. We start by presenting a comparison with established methods, and then empirically evaluate the performance of our algorithm on a variety of topologies, including random graphs, as well as regular lattices and real-world networks with a very large number of loops. Naturally, except for the calibration comparisons, our main focus will be the case of partial observations, where maximum likelihood based approaches are not applicable.

In all the tests below, unless stated otherwise, (i) parameters $\alpha_{ij}$ are sampled uniformly from $[0, 1]$; (ii) each cascade is generated independently from the IC model with limited $T$, varying from $4$ to $20$; (iii) the source of every cascade is a single, randomly chosen, node and (iv) hidden nodes are chosen uniformly at random. A non-uniform distribution of seeds and hidden nodes, and a use of multiple sources do not qualitatively change the reconstruction trends as long as sufficient statitics of activations is available, see a related discussion along with additional experiments in the Appendix, section \ref{appendix:2}. All points in the plots are averaged over multiple realisations of parameters $\alpha_{ij}^*$ and network structures -- where applicable, and the respective statistical confidence is presented through the error bars. We initialise the learning process with $\alpha_{ij} = 0.5 \, \, \forall {(ij)\in E}$, unless stated otherwise.

\subsection{Comparison with Established Methods}

We start with a few comparisons with existing methods to illustrate the basic properties of SLICER. DMPrec has been extensively compared to marginalized likelihood algorithms in terms of accuracy and running time in \cite{lokhov2016reconstructing}, and reconstructs model parameters from partial observations in an exact setting that we are considering here. Hence, it is natural to compare SLICER to this algorithm as a benchmark: from the discussion in the previous section, we expect to see an improvement in the algorithmic complexity. DMPrec has been originally introduced for a stochastic susceptible-infected model; in the Appendix, section \ref{appendix:3}, we adapt this algorithm to the case of the IC model considered here. For illustrating the accuracy of reconstruction obtained with SLICER, we benchmark our algorithm against the maximum likelihood approach; we only provide this ``calibration'' comparison in the case of full observability, where maximum likelihood is computationally tractable \cite{lokhov2016reconstructing}. Precise maximum likelihood based formulation is presented in the Appendix, Section \ref{appendix:4}. For the optimization of maximum likelihood, we use the optimization software Ipopt \cite{wachter2006implementation} within the Julia/JuMP modeling framework for mathematical optimization \cite{dunning2017jump}.

\begin{figure*}[ht]
    \includegraphics[width=0.46\textwidth]{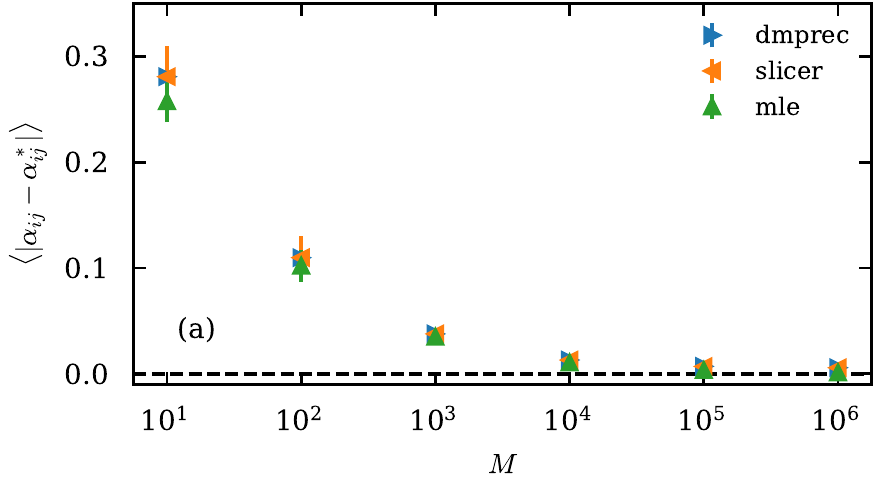}\hfill
    \includegraphics[width=0.46\textwidth]{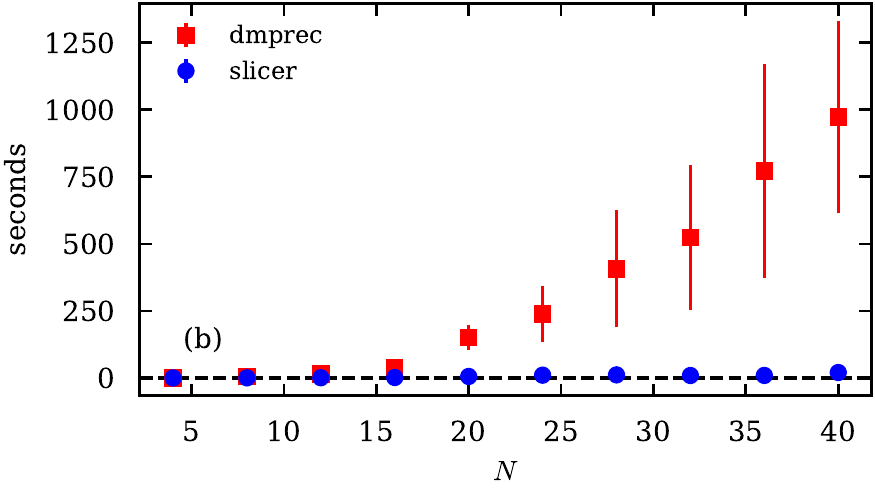}
    \caption{Comparison of accuracy and time complexity between SLICER, DMPrec and Maximum Likelihood estimators. (a) The average difference between inferred and real parameters $\alpha_{ij}$, as a function of number of available cascades for a 3-regular tree with $N=20$ nodes and cascades length $T=5$. Results averaged over five different sets of parameters. (b) Run-times comparisons of a single gradient step evaluation between SLICER and DMPrec as a function of network size. Results for 3-regular tree and $10^3$ cascades of length $T = 5$ averaged over 200 gradient steps evaluations and five different sets of parameters.}
    \label{fig:comparison}
\end{figure*}

Our experiments reveal that both SLICER and DMPrec produce identical results for parameter recovery, which is not surprising given that they optimize the same objective function. This phenomenon is illustrated in Fig. \ref{fig:comparison}(a), where both algorithms exhibit exactly the same accuracy. The main difference is the time needed to evaluate a single gradient step. As shown in Fig. \ref{fig:comparison}(b), a single-step gradient evaluation of DMPrec quickly reaches running time that are prohibitive for the application of the method to networks with more than several dozens of nodes. For this reason, we perform comparisons on relatively small networks with up to $N = 40$ nodes, consistent with the maximum size of test cases considered in \cite{lokhov2016reconstructing}. As we show below, SLICER maintains a low computational complexity, and can be easily scaled up to large networks; additional tests on computational complexity of SLICER as a function of problem parameters are presented in the Appendix, section \ref{appendix:1}.

Unlike maximum likelihood, SLICER optimizes the likelihood of marginal distributions (estimated with the DMP) only, which is the key to tractability in the case of partial observations. Still, as shown in Fig. \ref{fig:comparison}(a), it demonstrates a comparable accuracy to maximum likelihood on tree graphs, where DMP is exact, in the case of full observations.

\subsection{Learning Model Parameters}

\begin{figure*}
    \subfloat[$T = 5$]{\includegraphics[width=0.46\textwidth]{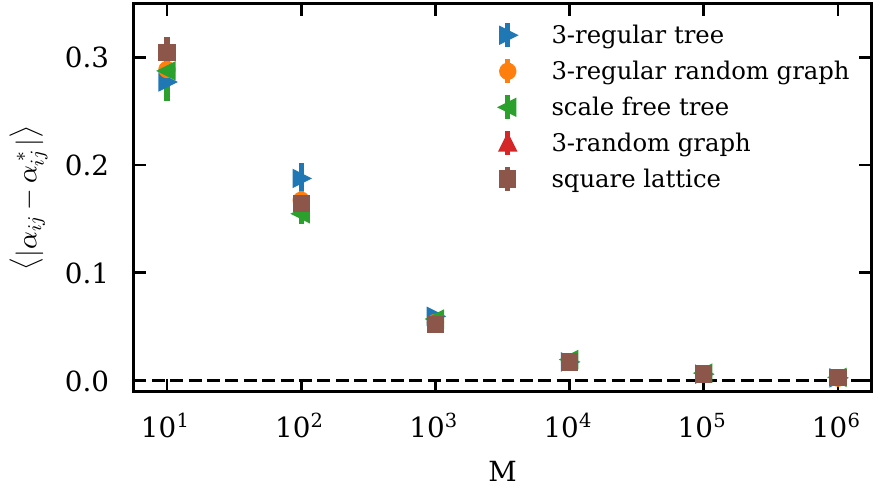}}\hfill
    \subfloat[$T = 10$]{\includegraphics[width=0.46\textwidth]{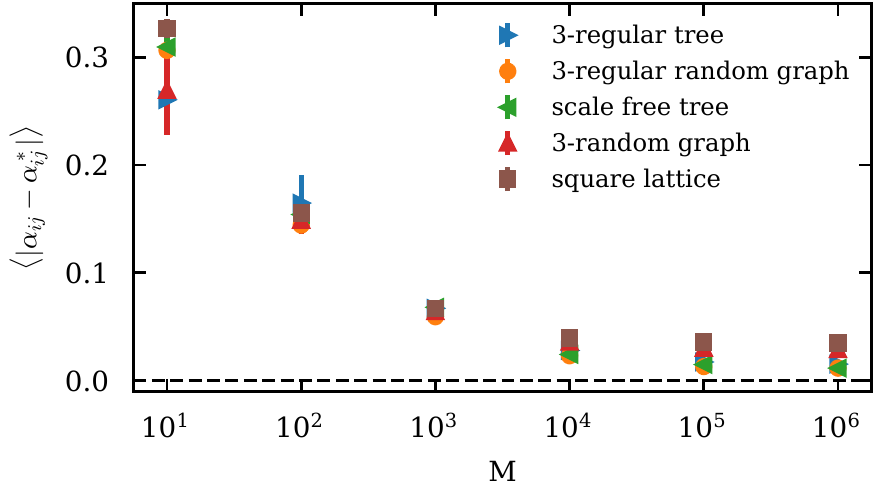}}
    \caption{Average difference between inferred $\alpha_{ij}$ and real $\alpha^*_{ij}$ parameters in $\ell_1$ norm, as a function of number of available cascades in the case of full observations of cascades, for two obervation horizons: (a) $T=5$ and (b) $T=10$. Each point is averaged over 5 different networks and 5 different sets of parameters $\alpha^*_{ij}$ (sampled uniformely from $[0, 1]$). All networks contain $N=100$ nodes.}
    \label{fig:full}
    \vspace{-0.05cm}
\end{figure*}

In order to get a flavor of typical reconstruction errors as a function of network topology, we calibrate our approach on the case of full observations. The results for this simplest setting, for different network types, are shown in Fig. \ref{fig:full}. In particular, we test error decay on a regular tree of degree three, a scale-free tree, a random regular graph of degree three, an Erd\"{o}s-R\'{e}nyi network with average degree three and a square lattice.
The apparent difference between left and right subplots of Fig.~\ref{fig:full} is a result of loops affecting the accuracy of DMP equations: When the observation time horizon is short, even though the graph is loopy, the spreading effectively occurs in a tree-like region where DMP equations are exact.
As $T$ increases, information spreads through loops that start to have an effect on the accuracy of predictions of marginals, and hence on the recovered parameters. For the adversarial case of a regular grid, the error saturates to a finite value for $T = 10$ even for a large number of samples. For parameter estimation task, this is expected due to the presence of systematic short loops.

\begin{figure*}[ht]
    \subfloat[3-regular tree, $T=5$]{\includegraphics[width=0.46\textwidth]{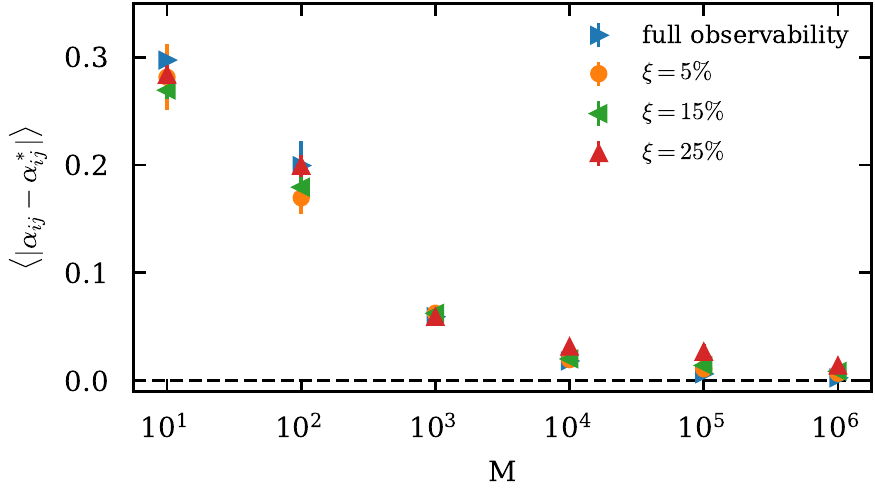}}\hfill
    \subfloat[3-random regular graph, $T=5$]{\includegraphics[width=0.46\textwidth]{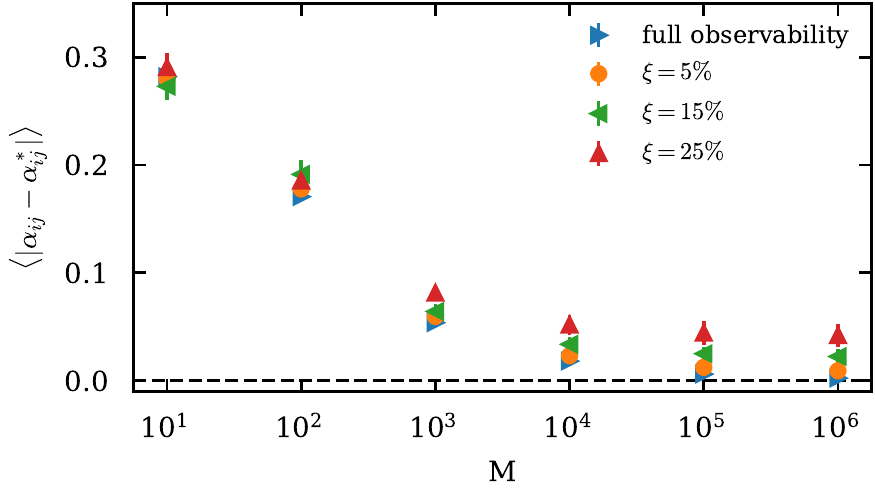}}\hfill
    \subfloat[3-regular tree, $T=10$]{\includegraphics[width=0.46\textwidth]{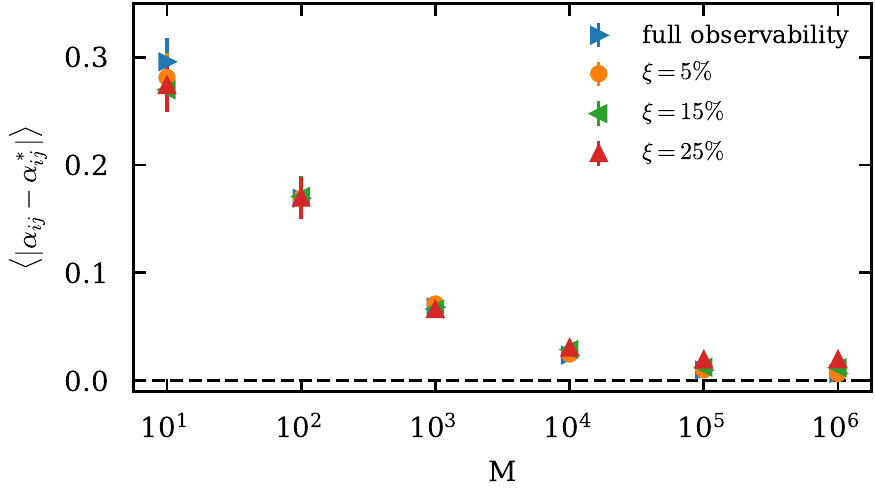}}\hfill
    \subfloat[3-random regular graph, $T=10$]{\includegraphics[width=0.46\textwidth]{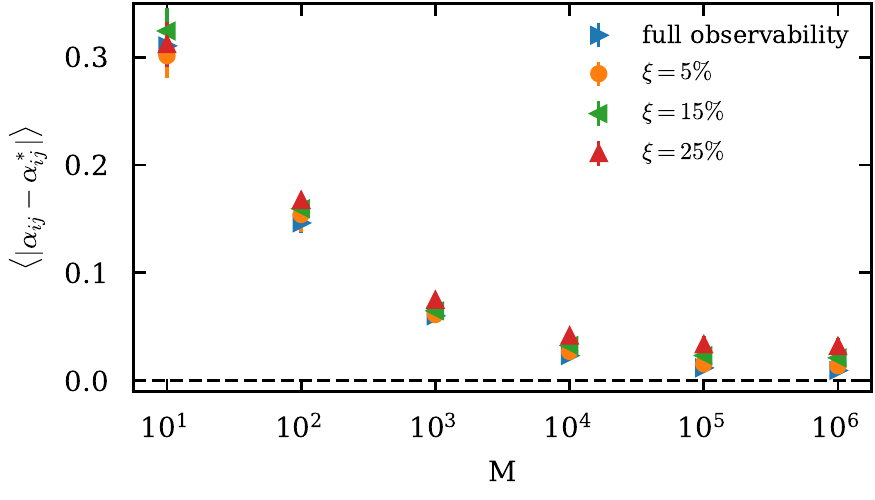}}
    \caption{Average difference between inferred $\alpha_{ij}$ and real $\alpha^*_{ij}$ parameters in $\ell_1$ norm, as a function of number of available cascades for different network types in the case where a fraction of nodes $\xi$ is unobserved. Each point is averaged over five different networks and five different sets of parameters $\alpha^*_{ij}$ (sampled from a uniform distribution between zero and one). All networks contain $N=100$ nodes. Unobserved nodes were picked at random. For tree graphs, edges adjacent to unobserved leaves were excluded.}
    \label{fig:partial}
\end{figure*}

Let us now consider a more challenging case of reconstructing parameters when only part of the information is available. To test the accuracy of our method, we randomly pick a fraction of nodes $\xi$ and hide their activation times throughout all cascades. Results for this setting are shown in Fig. \ref{fig:partial}. We test two types of networks, regular trees and regular random graphs, both for shorter, $T=5$, and longer, $T=10$, cascades. In a particular case of trees, we neglected the edges which connect the networks with unobserved leaves, since estimating their spreading parameters is impossible with any algorithm.
These results show that $\alpha_{ij}^*$ can be efficiently estimated to a good accuracy even with $\xi = 25\%$ of unobserved nodes, as long as enough data is available. Moreover, the reconstruction quality improves with the growing observation time for these tree-like cases.

\begin{figure*}[ht]
    \subfloat[Internet real-world network, $N=22,963$]{\includegraphics[width=0.46\textwidth]{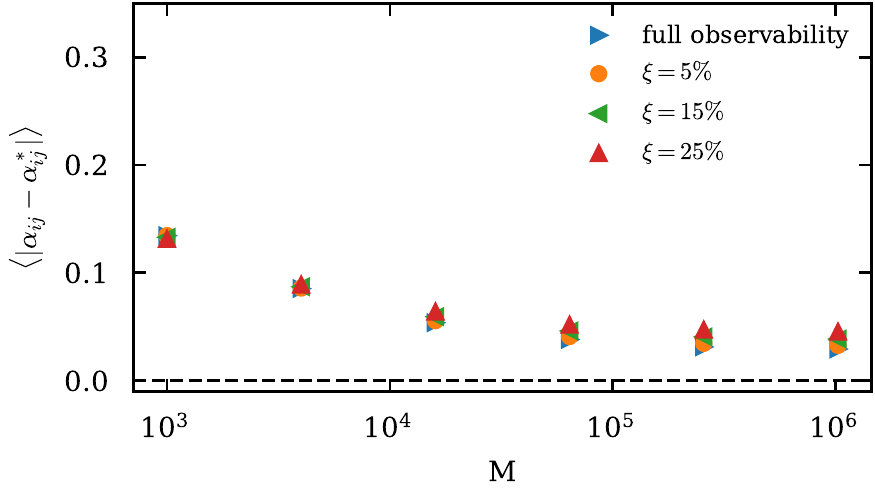}}\hfill
    \subfloat[Web-sk real-world network, $N=121,422$]{\includegraphics[width=0.46\textwidth]{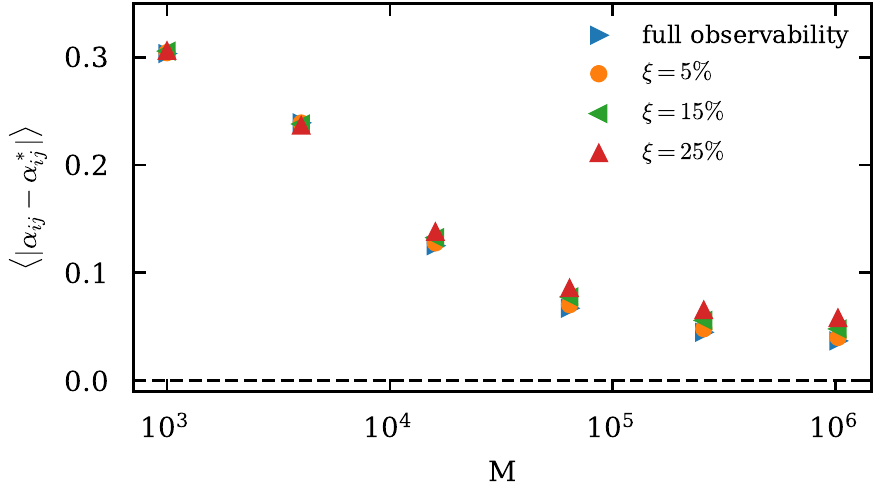}}
    \caption{Average difference between inferred $\alpha_{ij}$ and real $\alpha^*_{ij}$ parameters in $\ell_1$ norm, as a function of number of available cascades for two different empirical networks, where a fraction of nodes $\xi$ is unobserved. Unobserved nodes were picked at random.}
    \label{fig:empirical}
\end{figure*}

Finally, we test SLICER on loopy real-world networks.
We use two web networks for numerical tests: one representing the snapshot of the structure of the Internet at the level of autonomous systems \cite{rossi2015network}, and the other obtained by a web-crawler \cite{boldi2004ubicrawler}.
These networks are not only loopy ($46,873$ and $993,421$ triangles respectively), but also large ($N=22,963$ and $N=121,422$ nodes, $M=48,436$ and $M=334,419$ edges respectively), which allows us to test the scalability of SLICER.
We use a stochastic initial condition, where instead of a single source, the nodes activate at time $t=0$ randomly with $p^c_i(0) = \frac{10}{N}$ for each of the cascade. This means that each cascade starts with ten activated nodes on average.
In the learning procedure, the stochastic initial condition has an advantage that the DMP equations need to be run only once.
We chose the stochastic initial conditions for making sure that a large part of a network can be activated, thus providing information on model parameters.
Due to the presence of high-degree nodes, uniform choice of initial conditions leads to a rapid full activation of the network. We hence chose the degree-dependent transmission probabilities $\alpha^*_{ij}=[\max(k_i, k_j)]^{-\frac{1}{2}}$ that lead to a non-trivial dynamics, representing a trade-off between immediate activation of the full network, and a limited coverage of the graph. For this choice of parameters, $T=4$ turns out to be the time horizon beyond which DMP predicts the activation of the full network; we hence use the cascade data up to this time in our numerical experiments. In this test case, SLICER is initialized with small $\alpha_{ij}$.
Results are shown in Fig. \ref{fig:empirical}, and show a similar trend compared to the synthetic instances analyzed previously.
It is however important to note that the maximum considered data size of $10^6$ cascades represents a small statistics compared to the synthetic cases with $N=100$. Indeed, the ratio between the available data ($10^6$ cascades in both cases) and the number of parameters in real-world instances ($48,436$ and $334,419$ parameters respectively) is much lower than in synthetic cases ($150$ to $200$ parameters). Hence, we anticipate that a better accuracy can be achieved by collecting more data. Finally, structural particularities of realistic networks with interconnected hubs may render some edges non-identifiable in some regime of parameters. We did not filter our results based on the identifiability, and our error estimates may include random guesses for those links (identifiability is an interesting question but beyond the current scope).
Interestingly enough, despite a large number of short loops, it appears that partial observability has a small effect on the quality of the reconstructed parameters: reconstruction quality does not seem to drop much even for $\xi=25\%$. Overall, tests on real-world networks demonstrate that SLICER can scale to large networks with hundreds of thousands of nodes, a significant improvement compared to the DMPrec algorithm of \cite{lokhov2016reconstructing}. Learned model parameters even on loopy instances can then be used in the setting of ``inferning'' discussed next.

\section{Prediction of Marginal Probabilities}

As explained in the introduction, in many applications falling under the ``learning for inference'' setting \cite{heinemann2014inferning, bresler2020learning}, we are less interested in an accurate recovery of the underlying model parameters, but instead want to learn an effective model that succeeds in correctly predicting the marginal distributions that can subsequently be used for control, optimization, or learning tasks. Here, we show that our approach naturally recovers the best effective parameters that lead to a better approximation of the marginals in the case where DMP is used as an inference method due to its attractive computational complexity comparable to a single run of the Monte-Carlo simulation \cite{lokhov2019scalable}.

\begin{figure*}[ht]
    \includegraphics[width=0.46\textwidth]{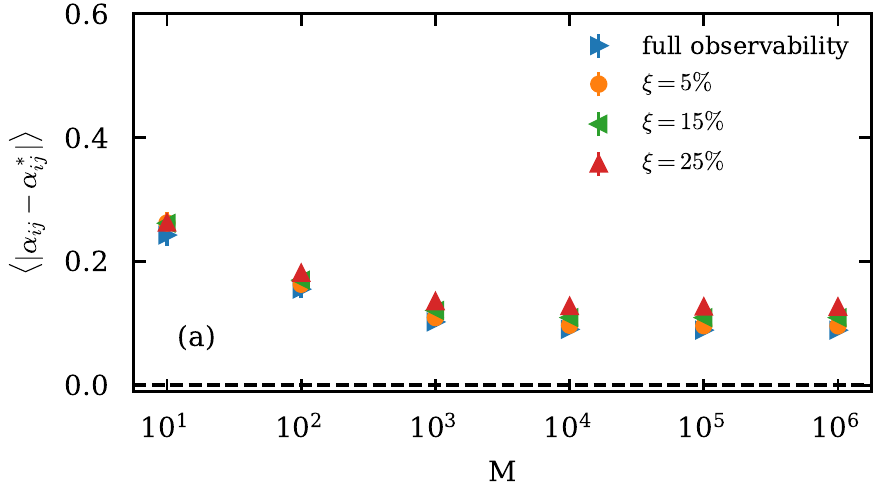}\hfill
    \includegraphics[width=0.46\textwidth]{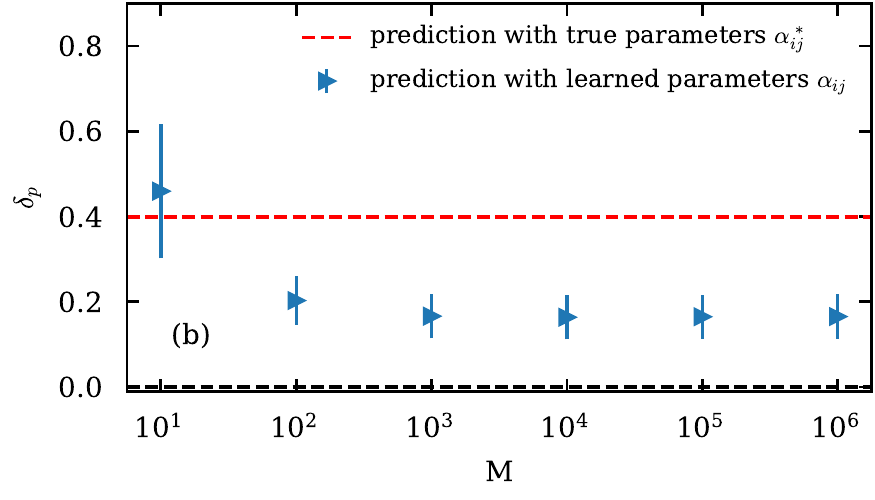}
    \vspace{-0.39cm}
    \caption{Illustration of the effective model concept, where our algorithm recovers the set of parameters that try to maximize the likelihood of marginal probabilities. Both panels correspond to an adversarial for DMP square lattice topology with many short loops, $N=100$ nodes, and cascades of length $T=20$.
    (a) Distance between true $\alpha^*_{ij}$ and estimated $\alpha_{ij}$ parameters.
    (b) Relative distance between ``ground-truth'' (estimated through Monte-Carlo) and estimated (via DMP) marginals. The red line represents the relative distance between real marginals and the ones obtained by using the ground-truth parameters $\alpha^{*}_{ij}$ within the DMP equations.}
    \label{fig:vs_marginal}
\end{figure*}

\subsection{Effective Model for Prediction of Marginals}

As already mentioned, the primary source of error in the DMP equations is related to the presence of loops. Their presence may result in overestimating parameters to account for DMP neglecting information on some spreading paths. To study this effect in more details, we focused on the most challenging case of a square lattice with many short loops, in an adversarial case of very long cascades with $T=20$. As shown in Fig. \ref{fig:vs_marginal}(a), this represents a challenging setting in terms of prediction of marginals using the DMP-based algorithm, which results in a finite error for parameter estimation even for a large number of cascades.
On the other hand, it is clear that when optimizing the objective function \eqref{eq:objective}, the algorithm attempts to maximize the likelihood of observed marginal probabilities. Hence, the recovered effective parameters should be the ones that attempt to minimize the distance between marginal probabilities estimated from the model, and those observed in the data. Fig. \ref{fig:vs_marginal}(b) shows relative $\ell_1$ distance between learned and ``ground-truth'' marginals
\begin{equation}
    \delta_p = \frac{\langle \vert p^s_i(t) - p^{s*}_i(t) \vert \rangle}{\langle p^{s*}_i(t) \rangle},
\end{equation}
where $p^s_i(t)$ are marginals estimated with DMP from the learned model; $p^{s*}_i(t)$ are the ``ground-truth'' values obtained from the total of $10^8$ Monte Carlo sampled cascades, $10^6$ per each source $s$; and averaging $\langle \cdot \rangle$ is done for all nodes, times, and sources. Fig. \ref{fig:vs_marginal}(b) demonstrates that DMP equations run with reconstructed parameters produce a better approximation of real marginal probabilities, compared to DMP predictions using the ground-truth parameters $\alpha_{ij}^*$.
This observation yields a promising avenue for building an effective model, which may not reflect the real spreading parameters, but allows for a more accurate prediction of the dynamics, which can subsequently be used for tasks such as influence maximization \cite{kempe2003maximizing} or control \cite{nowzari2016analysis}.

\begin{figure}[ht]
    \includegraphics[width=0.97\columnwidth]{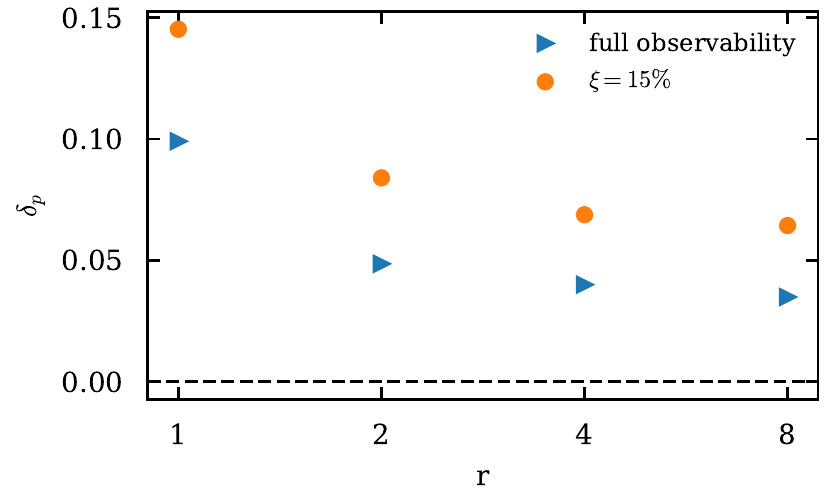}
    \vspace{-0.34cm}
    \caption{Improvement of the marginal prediction error using a multi-replica effective model as a function of the number of replicas for full $\xi = 0\%$ and partial $\xi = 15 \%$ observability.
    The results are obtained for a square lattice with $N = 100$ nodes and $M = 10^6$ cascades of length $T = 20$.}
    \label{fig:replicas}
    \vspace{-0.1cm}
\end{figure}

\begin{figure}[ht]
    \includegraphics[width=0.97\columnwidth]{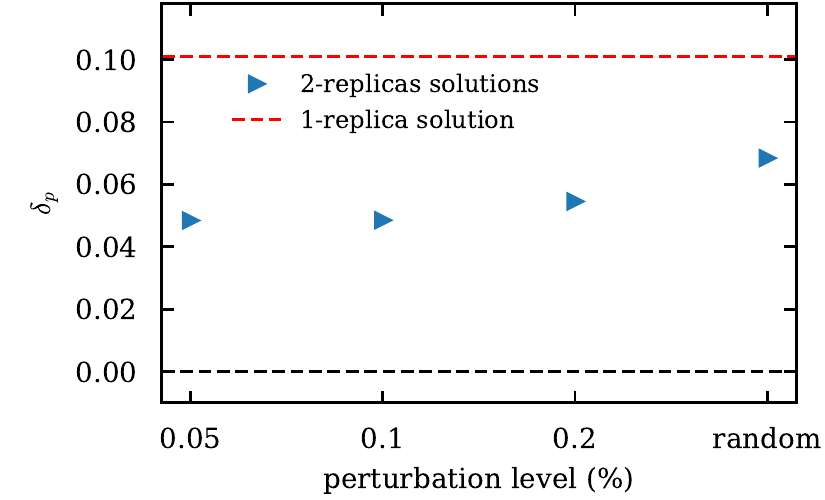}
    \vspace{-0.16cm}
    \caption{Dependence of the improvement in prediction of marginal probabilities by a 2-replica effective model compared to a single-replica model learned by SLICER as a function of perturbation of the previously obtained solution. The last point corresponds to a search of optimal parameters from scratch using random initialization.
    The results were obtained for a square lattice with $N = 100$ nodes and $M = 10^6$ cascades of length $T = 20$.}
    \label{fig:init}
    \vspace{-0.41cm}
\end{figure}

These results further confirm the need for coherence in the ``learning for inference'' approach. Inference tasks are typically computationally harder compared to their learning counterpart. An ideal estimation procedure would perfectly reconstruct model parameters, that may yield inaccurate predictions of the dynamics outcome due to intrinsic errors in approximate inference algorithms. Our analysis shows that a coherent use of an approximate inference method in both learning and inference task leads to an effective model that has a much stronger prediction power. This observation is in agreement with previous work in learning of graphical models \cite{wainwright2006estimating, heinemann2014inferning} that showed that an inconsistent parameter estimation that takes approximate inference (such as belief propagation) into account is provably beneficial for the predictive task. This concept further motivates the extension of the search for effective models outside of the original class (for instance, allowing for an extended set of parameters), as long as the goal consists in an accurate prediction of the dynamics.

\subsection{Extension to Multi-Replica Effective Model}

Here, we capitalize on the concept of effective models for inference and develop a novel procedure for learning a mixture of spreading models on several replicas of the graph which further improves prediction quality of marginal probabilities.
Our proposed approach for introducing a mixture of replicas is motivated by the fact that DMP on a given graph instance is too constrained. Indeed, one may hope to adjust model parameters to compensate for the approximation error. However, in the context of other approximate inference algorithms such as belief propagation, it has been shown that some sets of valid marginals cannot be reached by any parameters or any learning procedure \cite{pitkow2011learning}; a similar picture is likely to apply to DMP on a given network. One could naturally try to look at extended graph topologies that could potentially bring some flexibility, however in the context of dynamic models considered here, additional edges do not benefit the effective model constructions since they are likely inconsistent with the observed cascades. However, several copies of the original topology are always consistent with the data, while they may compensate for errors in marginal prediction by adjusting model parameters in each replica. Notice that introduction of a mixture of replicas in our approach is a mean to increase the expressiveness of the inference model and does not mean that the cascade data is actually produced by the mixture of graphs \cite{hoffmann2020learning}.

To test this idea, we introduce several replicas of the model with independent parameters, and where the predicted marginal is given by a uniform mixture over these replicated models. Formally, we consider the objective:
\begin{equation}
    \mathcal{O}^{\text{mixture}} = \sum_{s \in S} \sum_{i \in O} \sum_{\tau_i^s} m^{\tau^s_i} \log \left( \frac{1}{|R|} \sum_{r \in R} \mu^s_{i, r}(\tau^s_i) \right),
    \label{eq:replicas}
\end{equation}
where $R$ is the set of replicas, and $\mu^s_{i, r}(t)$ is the marginal probability of node $i$ being activated at time $t$ for a cascade with source $s$ in replica model $r$. SLICER can be straightforwardly generalized to the objective function \eqref{eq:replicas} with a mixture of replica models. Thus obtained effective model is more expressive since the number of parameters is increased from $|E|$ to $|E| \cdot |R|$; the computational complexity of the algorithm for this case changes to $O(|R| |E| T |S|)$. We propose the following scheme for a systematic improvement of the prediction of marginal probabilities by introducing more replicas compared to the solution obtained at the previous step, where the original effective model obtained by SLICER represents a single-replica case. It consists in finding a solution for an effective model with an increased number of replicas by running SLICER initialized with a slightly perturbed solution (to break symmetry between replicas), previously obtained for a model with a smaller number of replicas. The results of this procedure are given in Fig. \ref{fig:replicas} for the case of a square lattice with long, $T = 20$, observation time. We see that subsequent introduction of replicas progressively and substantially improves the prediction quality of marginal probabilities, both in the case of full and partial observations. Fig. \ref{fig:init} shows that small perturbations from previously obtained solutions produce better results and guarantee improved prediction compared to a search for optimal parameters from scratch. A more detailed discussion of this phenomenon along with supporting experiments are given in the Appendix, section \ref{appendix:5}. This novel effective mixture model construction thus allows one to systematically overcome structural constraints of the DMP equations, and produce a model with higher-quality predictions.

\section{Conclusions}

Incomplete information about the dynamics significantly complicates the tasks of parameter estimation and learning of a model for generating accurate predictions.
In this paper, we propose a novel approximate learning algorithm, SLICER, based on a dynamic message-passing approach. SLICER achieves a significantly better scalability compared to existing approximate methods. Moreover, even in a loopy regime which is adversarial for message-passing, our method explores the space of parameters in search of an effective model that achieves a better prediction of the dynamics. Finally, we propose a novel procedure for estimating a multi-replica effective model that systematically further improves prediction quality of marginal probabilities. We anticipate that scalability and prediction qualities of our algorithm will be useful for real-world applications where large amounts of incomplete data is available and a need of reliable scenario testing is essential.
Although our study used the standard discrete-time IC model for illustration, DMP can be generalised to more complex dynamical models \cite{lokhov2015dynamic}, as well as to models with time varying parameters $\alpha_{ij}(t)$, adaptive spreading settings, continuous time dynamics or temporal networks.
It would be also insightful to extend our approach to the point processes on networks \cite{upadhyay2018deep,farajtabar2017coevolve}.
Finally, we believe that our approach will be helpful in the task of optimal resource allocation for both monitoring and containment of spreading processes.

\subsection*{Implementation of the Algorithms}
A full implementation of our algorithms is available at \cite{wilinski2021code}.

\subsection*{Acknowldegments}

Authors acknowledge support from the Laboratory Directed Research and Development program of Los Alamos National Laboratory under projects numbers 20200121ER and 20210529CR.

\bibliography{references}
\bibliographystyle{icml2021}

\onecolumn

\appendix
\def\thesection{A\arabic{section}}
\renewcommand{\thefigure}{A\arabic{figure}}
\renewcommand{\theequation}{A\arabic{equation}}
\renewcommand{\thetable}{A\arabic{table}}

\setcounter{equation}{0}
\setcounter{figure}{0}
\setcounter{table}{0}

{\Large \bf Appendix}

\section{Details of SLICER implementation}
\label{appendix:1}

\subsection{Full derivation of the algorithm}
\label{sec:framework}

For consistency, let us re-state the DMP equations for the IC model used in the Main Text, Eqs. (\ref{eq:marginal})-(\ref{eq:message}):
\begin{equation}
    p^c_i(t) = 1 - \big( 1 - \bar{p}^c_i \big) \prod_{k \in \partial i} \Big(1 - \alpha_{ki} \cdot p^c_{k \rightarrow i}(t - 1) \Big),
    \label{eq:marginal_a}
\end{equation}
\begin{equation}
    p^c_{j \rightarrow i}(t) = 1 - \big( 1 - \bar{p}^c_j \big) \prod_{k \in \partial j \setminus i} \Big(1 - \alpha_{kj} \cdot p^c_{k \rightarrow j}(t - 1) \Big),
    \label{eq:message_a}
\end{equation}
which allow to compute $p^c_i(t)$ - the probability of node $i$ to be active at time $t$, under the initial condition for the cascade $c$. As specified in the Main Text, assuming that each cascade has a single node source, we can write the objective function as:
\begin{equation}
    \begin{split}
        \mathcal{O} &= \sum_{s \in S} \sum_{i \in O} \sum_{\tau_i^s} m^{\tau^s_i} \log \mu^s_i(\tau^s_i),
    \end{split}
\end{equation}
where $m^{\tau^s_i}$ is the number of times when node $i$ was activated at time $\tau^s_i$ for all cascades in the data that start at node $s$, and $\mu^s_i(\tau^c_i)$ is the marginal probability that node $i$ for an initial source $s$ is activated precisely at time $\tau^s_i$. These marginals can be approximately expressed through the DMP equations, which results in the following expression:
\begin{equation}
    \mathcal{O} = \sum_{s \in S} \sum_{i \in O} \sum_{\tau^s_i} m^{\tau^s_i} \log\Big( p^s_i(\tau^s_i) \cdot \mathds{1}_{(\tau^s_i < T)} - p^s_i(\tau^s_i - 1) \cdot \mathds{1}_{(\tau^s_i > 0)} + \mathds{1}_{(\tau^s_i = T)} \Big),
\end{equation}
where $T$ is the maximum cascade length.

The constraints are given by DMP equations re-weighted by Lagrange multipliers:
\begin{equation}
    \begin{split}
        \mathcal{C} &= \sum_{s \in S} \sum_{t=0}^{T-1} \sum_{i \in V} \lambda^s_i(t+1) \Bigg( p^s_i(t+1) - 1 + \big( 1 - \bar{p}^s_i \big) \prod_{k \in \partial i} \bigg(1 - \alpha_{ki} \cdot p^s_{k \rightarrow i}(t) \bigg) \Bigg) \\
        &+\sum_{s \in S} \sum_{t=0}^{T-1} \sum_{(i,j) \in E} \lambda^s_{i \rightarrow j}(t+1) \Bigg( p^s_{i \rightarrow j}(t+1) - 1 + \big( 1 - \bar{p}^s_i \big) \prod_{k \in \partial i \setminus j} \bigg(1 - \alpha_{ki} \cdot p^s_{k \rightarrow i}(t) \bigg) \Bigg),
    \end{split}
\end{equation}
where $\lambda^s_i(t)$ and $\lambda^s_{i \rightarrow j}(t)$ are Lagrange multipliers corresponding respectively to marginals from (\ref{eq:marginal_a}) and messages from (\ref{eq:message_a}).

The resulting expressions for all the quantities can be found by differentiating the Lagrangian. Differentiation over marginals yields expressions for $\lambda^s_i(t)$ multipliers for all sources $s$ and times $t$:
\begin{equation}
\begin{split}
    \frac{\partial \mathcal{L}}{\partial p^s_i(t)} &= \lambda^s_i(t) + \sum_{\tau_i^s} \frac{m^{\tau^s_i} \cdot \mathds{1}_{(t=\tau^s_i)} \cdot \mathds{1}_{(\tau^s_i < T)}}{p^s_i(\tau^s_i) - p^s_i(\tau^s_i - 1) \cdot \mathds{1}_{(\tau^s_i > 0)}} \\
    &+ \sum_{\tau_i^s} \frac{m^{\tau^s_i} \cdot \mathds{1}_{(t=\tau^s_i-1)} \cdot \mathds{1}_{(\tau^s_i > 0)}}{p^s_i(\tau^s_i - 1) - p^s_i(\tau^s_i) \cdot \mathds{1}_{(\tau^s_i < T)} - \mathds{1}_{(\tau^s_i = T)}}.
\end{split}
\end{equation}
One direct consequence of this is that $\lambda^s_i(t) = 0 \, \forall_{t \notin \{ \tau^s_i, \tau^s_i - 1 \}}$. The $\lambda^s_{i \rightarrow j}(t)$ multipliers can be obtained from the derivatives over messages.
For $t < T$ we get
\begin{equation}
    \begin{split}
        \frac{\partial \mathcal{L}}{\partial p^s_{i \rightarrow j}(t)} &= \lambda^s_{i \rightarrow j}(t) - \lambda^s_j(t+1) \, \alpha_{ij} \, \big( 1 - \bar{p}^s_j \big) \prod_{m \in \partial j \setminus i} \Big(1 - \alpha_{mj} \cdot p^s_{m \rightarrow j}(t) \Big) \\
        &- \sum_{k \in \partial j \setminus i} \lambda^s_{j \rightarrow k}(t+1) \, \alpha_{ij} \, \big( 1 - \bar{p}^s_j \big) \prod_{m \in \partial j \setminus \{ i, k \}} \Big(1 - \alpha_{mj} \cdot p^s_{m \rightarrow j}(t) \Big).
    \end{split}
    \label{eq:Lagrange_multipliers1}
\end{equation}
For $t = T$, the expression simplifies to
\begin{equation}
    \frac{\partial \mathcal{L}}{\partial p^s_{i \rightarrow j}(T)} = \lambda^s_{i \rightarrow j}(T),
    \label{eq:Lagrange_multipliers2}
\end{equation}
which allows to calculate all $\lambda^s_{i \rightarrow j}(t)$ in an inductive manner, starting from $T$ and using $\lambda^s_{i \rightarrow j}(t+1)$ to compute multipliers for $t<T$. Finally, derivatives over parameters $\alpha_{ij}$ read
\begin{equation}
    \begin{split}
        \frac{\partial \mathcal{L}}{\partial \alpha_{ij}} = &- \sum_{s \in S} \sum_{t=0}^{T-1} \lambda^s_i(t+1) \, p^s_{j \rightarrow i}(t) \, \big( 1 - \bar{p}^s_i \big) \prod_{m \in \partial i \setminus j} \Big(1 - \alpha_{mi} \cdot p^s_{m \rightarrow i}(t) \Big) \\
        &-\sum_{s \in S} \sum_{t=0}^{T-1} \lambda^s_j(t+1) \, p^s_{i \rightarrow j}(t) \, \big( 1 - \bar{p}^s_j \big) \prod_{m \in \partial j \setminus i} \Big(1 - \alpha_{mj} \cdot p^s_{m \rightarrow j}(t) \Big) \\
        &-\sum_{s \in S} \sum_{t=0}^{T-1} \sum_{k \in \partial i \setminus j} \lambda^s_{i \rightarrow k}(t+1) \, p^s_{j \rightarrow i}(t) \, \big( 1 - \bar{p}^s_i \big) \prod_{m \in \partial i \setminus \{ j, k \}} \Big(1 - \alpha_{mi} \cdot p^s_{m \rightarrow i}(t) \Big) \\
        &-\sum_{s \in S} \sum_{t=0}^{T-1} \sum_{k \in \partial j \setminus i} \lambda^s_{j \rightarrow k}(t+1) \, p^s_{i \rightarrow j}(t) \, \big( 1 - \bar{p}^s_j \big) \prod_{m \in \partial j \setminus \{ i, k \}} \Big(1 - \alpha_{mj} \cdot p^s_{m \rightarrow j}(t) \Big),
    \end{split}
    \label{eq:alpha_a}
\end{equation}
which can further be simplified for $\alpha_{ij} \neq 0$:
\begin{equation}
    \frac{\partial \mathcal{L}}{\partial \alpha_{ij}} = \frac{-1}{\alpha_{ij}} \sum_{s \in S} \sum_{t=0}^{T-1} \left(\lambda^s_{i \rightarrow j}(t) \, p^s_{i \rightarrow j}(t) + \lambda^s_{j \rightarrow i}(t) \, p^s_{j \rightarrow i}(t)\right).
\end{equation}
Now, we can use gradient components (\ref{eq:alpha_a}) to update parameters $\alpha_{ij}$:
\begin{equation}
    \alpha_{ij} \longleftarrow \alpha_{ij} + \varepsilon \cdot \frac{\partial \mathcal{L}}{\partial \alpha_{ij}},
    \label{eq:step_a}
\end{equation}
where $\varepsilon$ is a learning rate. The way we specify this learning rate is described in section \ref{sec:rate}.

\subsection{Choice of the learning rate}
\label{sec:rate}

\begin{figure*}[!h]
    \centering
    \includegraphics[width=0.75\columnwidth]{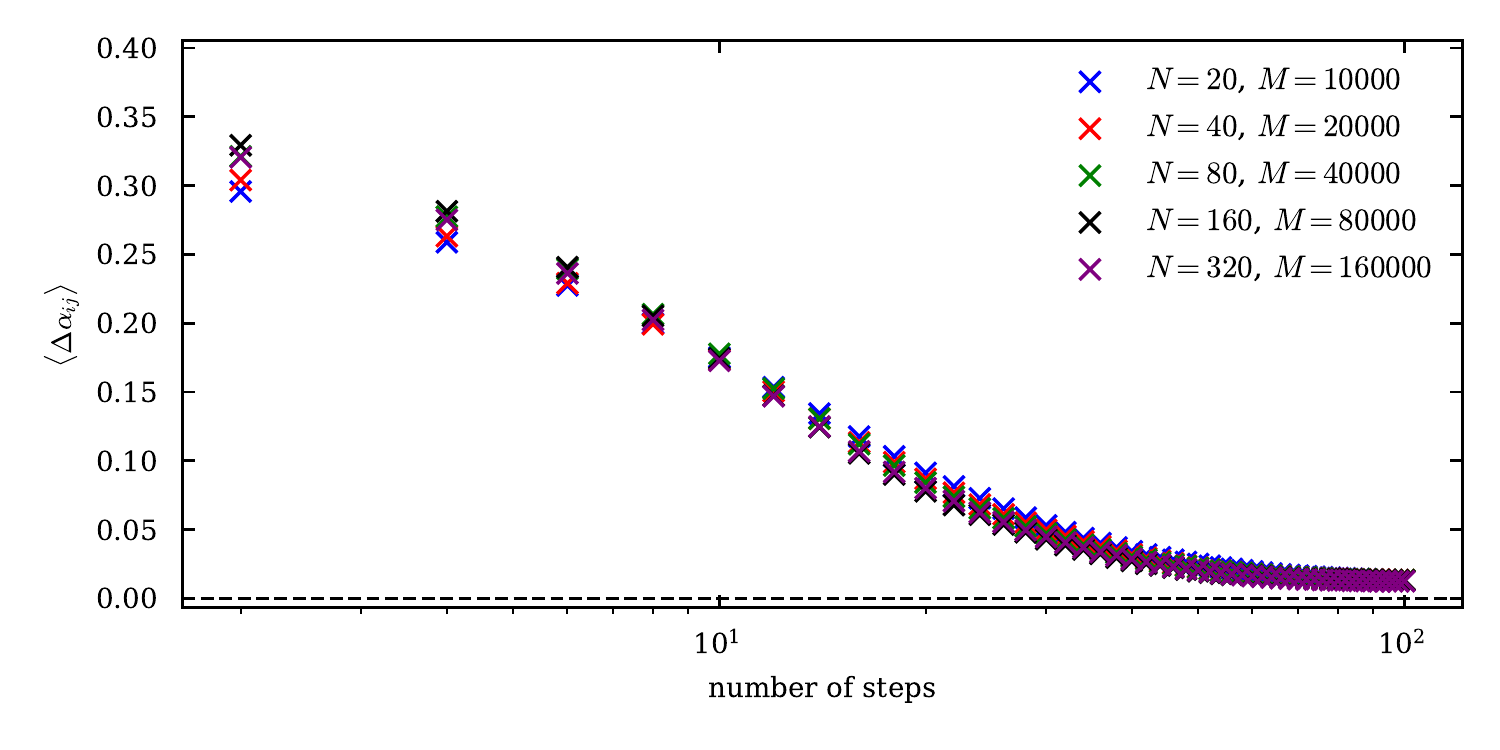}
    \caption{Convergence of the algorithm as a function of number of gradient descent steps, measured by the change of $\alpha_{ij}$ after each consecutive step.
    The simulations were done with cascades of length $T=10$.
    All the points were obtained using the learning rate $\varepsilon = c \frac{N}{M T}$ in the equation (\ref{eq:step_a}).}
    \label{fig:sup_convergence}
\end{figure*}

All of the numerical results presented in the paper were obtained with the learning rate $\varepsilon = c \frac{N}{M T}$ in equation (\ref{eq:step_a}), where $N$ is the number of nodes, $M$ is the number of cascades, $T$ is the length of cascades, and $c$ is a small constant (the same for all networks) ensuring that the step is not too large (in our experiments, we used $c = 1/80$). This normalisation of the gradient step is convenient because it results in very similar number of steps until global convergence across different network sizes for similar level of available information. This effect is demonstrated in Fig.~\ref{fig:sup_convergence} in the case of random three regular graph with varying size and accordingly re-scaled number of cascades. As a consequence of this observation, it is easy to estimate the running time of the algorithm based on the running time of a single iteration. In the next section, we show the empirical scalability per iteration step of the proposed algorithm as a function of problem parameters.

\subsection{Empirical algorithmic complexity}

\begin{figure*}[!h]
    \centering
    \includegraphics[width=0.75\columnwidth]{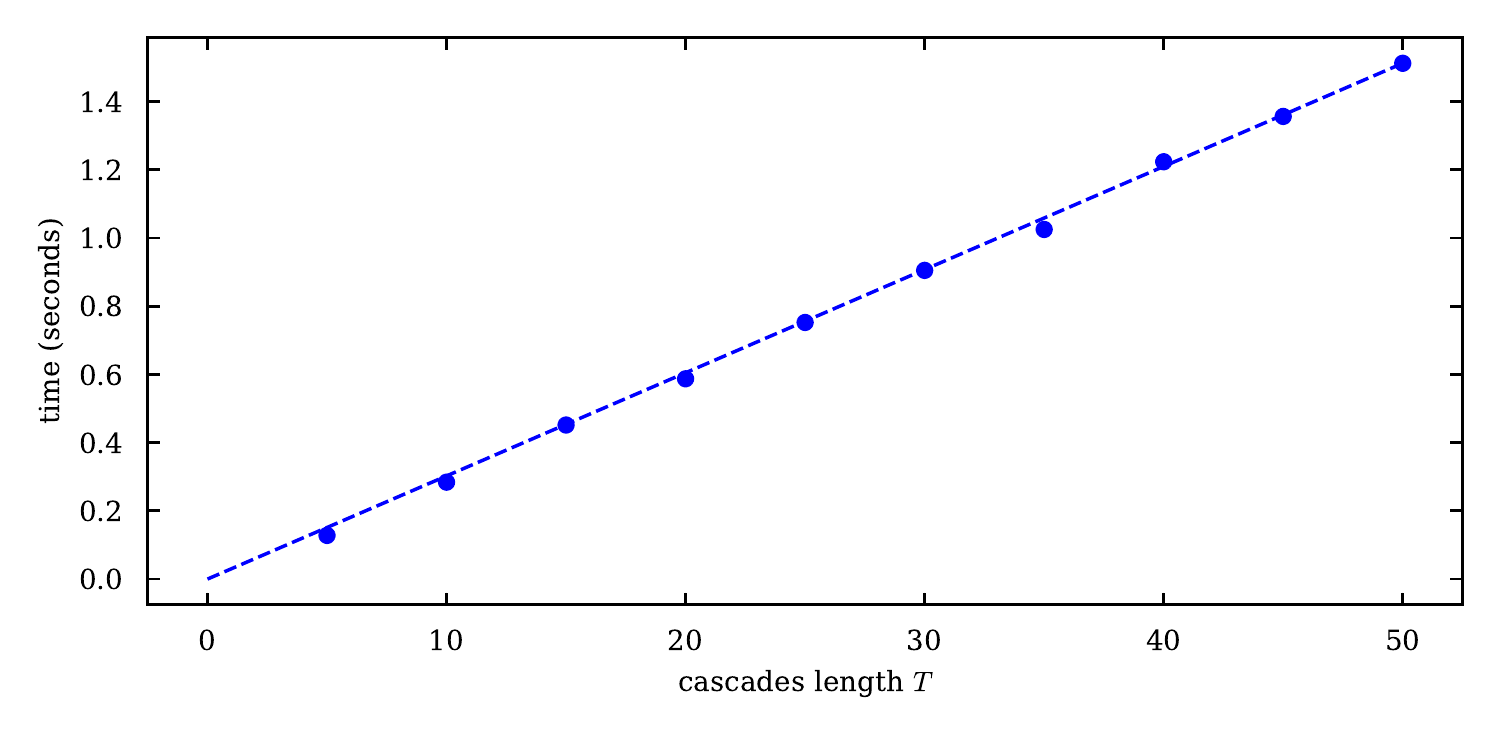}
    \caption{Averaged computational times for a single step of optimisation, as a function of cascade length $T$. Simulations were made using $M=100$ cascades on a random three regular graph with $N=100$ nodes. Each point represents an average over five different realisations of the network. The reference dashed line is the best linear fit with a zero intercept.}
    \label{fig:sup_length}
\end{figure*}

\begin{figure*}[!h]
    \centering
    \includegraphics[width=0.75\columnwidth]{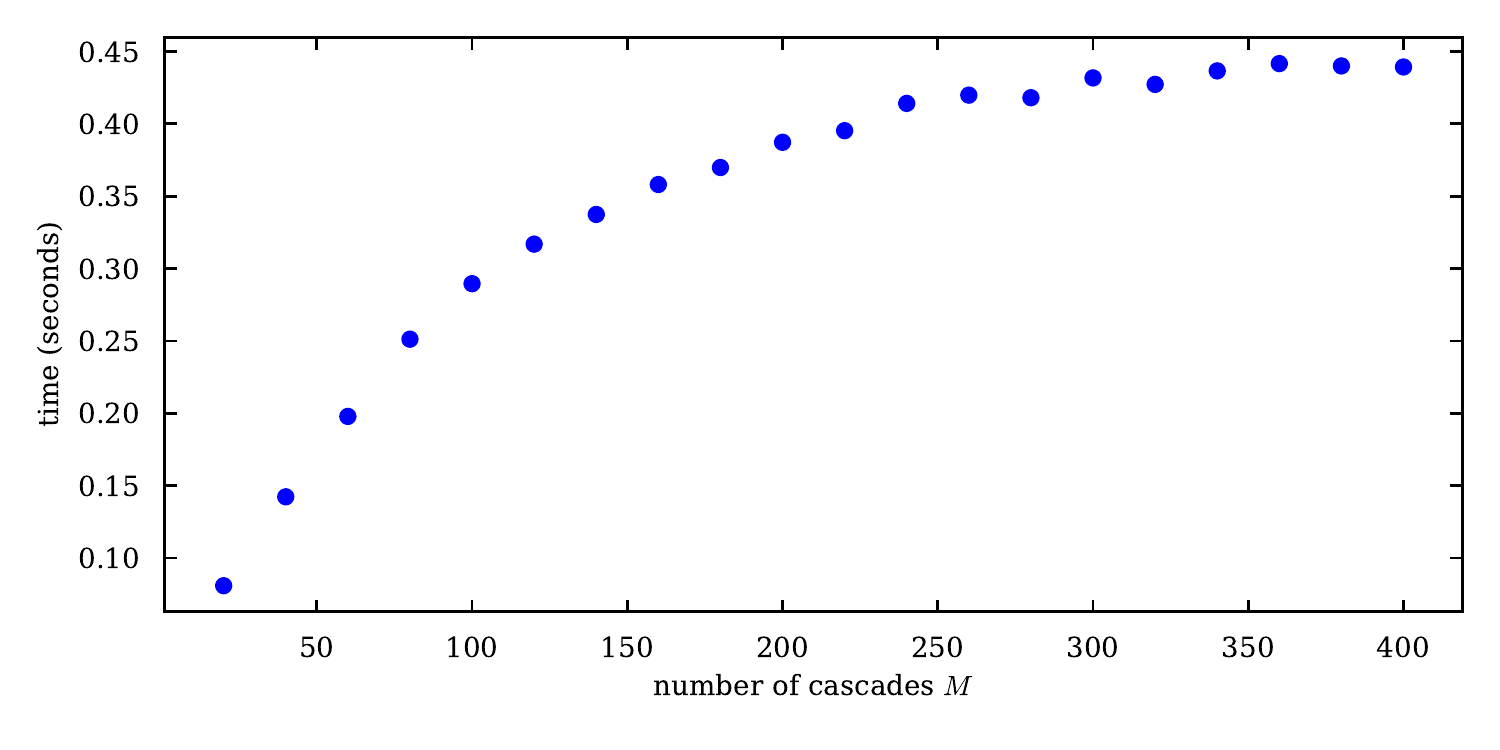}
    \caption{Averaged computational times for a single step of optimisation, as a function of the number of available cascades $M$. Simulations were made using cascades of $T=10$ length on a random three regular graph with $N=100$ nodes. Each point represents an average over five different realisations of the network.}
    \label{fig:sup_cascades}
\end{figure*}

\begin{figure*}[!h]
    \centering
    \includegraphics[width=0.75\columnwidth]{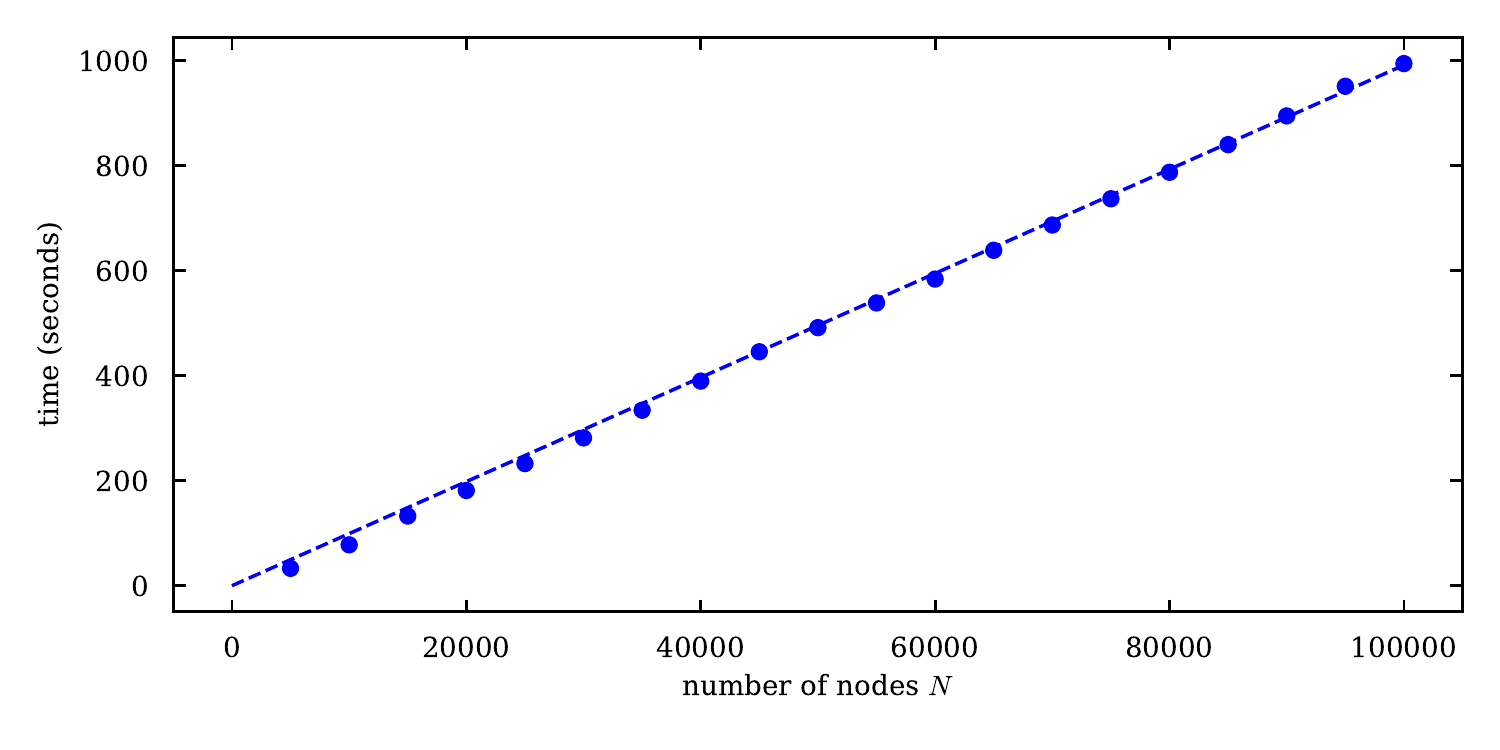}
    \caption{Averaged computational time for a single step of optimisation, as a function of the network size $N$. Simulations were made using $M=100$ cascades of $T=10$ length on a random three regular graphs of different sizes. Each point represents an average over five different realisations of the network. The reference dashed line is the best linear fit with a zero intercept.}
    \label{fig:sup_size}
\end{figure*}

As discussed in the Main Text, the overall computational complexity of SLICER is $O(|E| T |S|)$. In this section, we check the empirical scalability of the algorithm. Fig.~ \ref{fig:sup_length} shows the linearly growing computational time required to perform one step of the optimisation process as a function of the cascade length $T$. A more interesting dependence on the number of cascades is shown in Fig.~ \ref{fig:sup_cascades}. Linear until the number of cascades reaches the order of the network size, the optimization time starts to plateau for larger $M$ once the number of classes saturates to $|S|=N$ (we only use single node seeds in this experiment), which is consistent with the complexity analysis above. Note that for stochastic initial conditions $|S|=1$. Finally, the linear scaling for different network sizes is presented in Fig.~\ref{fig:sup_size}.

\subsection{Code and implementation efficiency}

The code used to obtain all the results in the paper is available at:
\url{https://github.com/mateuszwilinski/dynamic-message-passing}. We used a straightforward implementation of the proposed algorithm, where the DMP equations are run for all nodes and edges in the graph. This implementation choice was made so that it is applicable to general initial conditions that include arbitrary initial probability distributions factorized over nodes of the graph \cite{lokhov2019scalable}. However, for the single-source cascades considered in this paper for simplicity, it is clear that on very large networks one could take advantage of the finite $T$ and restrict the dynamics both in the primal and in the dual space to the part of the network that is reachable by the dynamics. Due to the linear scaling with the system size our learning scheme can be run on networks with the size on the order of millions of nodes. At these sizes the implementation could be further improved by a more efficient memory usage. Finally, notice that calculations for different cascades can be paralleled, where additionally an asynchronous stochastic version of the gradient descent can be used. 

\section{Further experiments on the experimental setting}
\label{appendix:2}

\subsection{Effect of initial number of activated nodes}

\begin{figure}[ht]
    \centering
    \includegraphics[width=0.75\columnwidth]{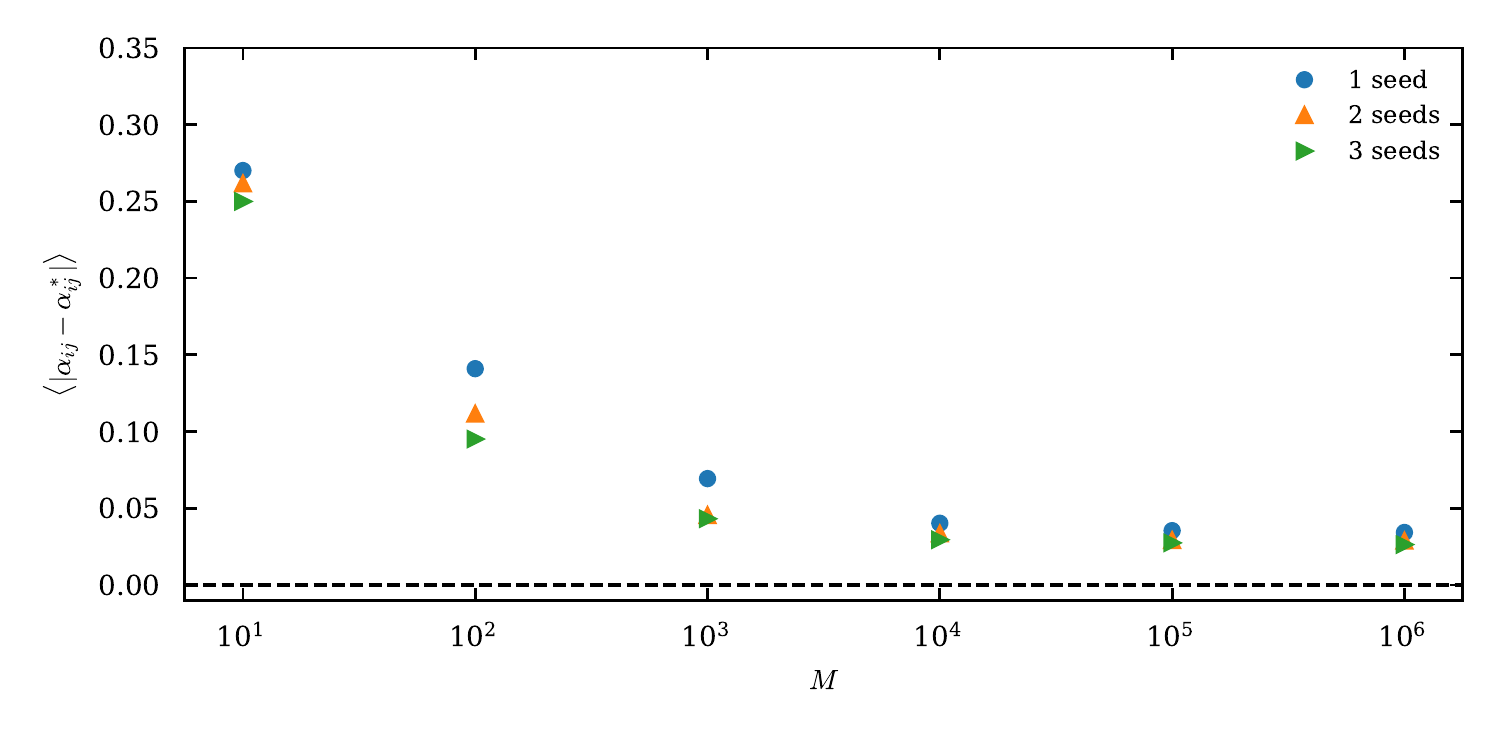}
    \caption{Results for Erd\"os-R\'enyi network with $N=100$ nodes, average degree of $3$, cascades length $T=10$, uniformly distributed parameters $\alpha_{ij}$ and different number of initially activated nodes.
    Seed size has little impact on reconstruction quality if single-source statistics has big network coverage.}
    \vspace{-0.3cm}
    \label{fig:seed}
\end{figure}

In the paper, we assume that cascades are random, and each one can start from any randomly-selected node.
To address the influence of initial conditions, we ran an experiment that shows reconstruction error as a function of the seed set size.
Results in Fig. \ref{fig:seed}, on a heterogeneous Erd\"os-R\'enyi network, show that increase in the seed size has a minor impact on the error if all nodes are activated a sufficient number of times, even if this happens in small-outbreak cascades with different sources.

\subsection{Effects of network topology and placement of the missing nodes in the network}

In the paper, we deliberately presented synthetic instances mostly with regular degree (grids and random graphs) to eliminate potential heterogeneity on the distribution of hidden nodes.
Nevertheless, a question on the influence of hidden node distribution is still interesting.
We ran additional experiments on a heterogeneous Erd\"os-R\'enyi network, comparing results for hidden node distributions targeting primarily high- or low-degree nodes with the random case, and otherwise in the same setting as throughout the paper.
The results in Table~\ref{tab:my_label} show better accuracy for hidden high-degree nodes due to dynamics going through them more frequently and hence accumulating more statics.

\begin{table}[H]
    \centering
    \begin{tabular}{c||c|c|c}
    hidden location & high-degree & random & low-degree \\
      \hline
    average $\ell_1$ error & 0.032   &  0.039 & 0.047
    \end{tabular}
    \caption{Average reconstruction error for the parameters as a function of distribution of hidden nodes location on an Erd\"os-R\'enyi network with $N=100$, $T=10$, $M=10^5$, and $\xi = 15\%$.}
    \label{tab:my_label}
\end{table}

\section{DMPrec derivation for the IC model}
\label{appendix:3}

The idea behind DMPrec is to compute the derivative over the DMP free energy of the system. The part of the free energy related to individual node $i$ can be written as:
\begin{equation}
    f^i_{\text{DMP}} = - \sum_c \log \mu^c_i(\tau_i^c),
\end{equation}
where marginal probability $\mu^c_i(\tau_i^c)$ of node $i$ being activated at time $t$ during cascade $c$ can be computed using marginals from DMP equations (\ref{eq:marginal_a}):
\begin{equation}
    \mu^c_i(t) = p^c_i(t) \mathds{1}_{(t < T)} - p^c_i(t-1) \mathds{1}_{(t > 0)} + \mathds{1}_{(t = T)}.
\end{equation}
The derivative of the DMP free energy over the spreading probabilities is as follows:
\begin{equation}
    \frac{\partial f^i_{\text{DMP}}}{\partial \alpha_{rs}} = - \sum_c \frac{1}{\mu^c_i(\tau_i^c)} \frac{\partial \mu^c_i(\tau_i^c)}{\partial \alpha_{rs}}.
\end{equation}
For cascades originating from a single source, we can further write:
\begin{equation}
    \frac{\partial f^i_{\text{DMP}}}{\partial \alpha_{rs}} = - \sum_s \sum_{\tau_i^s} \frac{m^{\tau_i^s}}{\mu^s_i(\tau_i^s)} \frac{\partial \mu^s_i(\tau_i^s)}{\partial \alpha_{rs}}.
    \label{eq:derivative}
\end{equation}
Let us now denote:
\begin{equation}
    \frac{\partial p^c_{k \rightarrow i}(t)}{\partial \alpha_{rs}} = \phi^{rs}_{k \rightarrow i}(t).
\end{equation}
Note that $\phi^{rs}_{k \rightarrow i}(0) = 0$.
We can use that and compute values for all other times in an iterative way, using:
\begin{equation}
    \phi^{rs}_{k \rightarrow i}(t) = \big( 1 - \bar{p}^c_k \big) \sum_{l \in \partial k \setminus i} \Big( \alpha_{lk} \cdot \phi^{rs}_{l \rightarrow k}(t-1) + p^c_{l \rightarrow k}(t-1) \mathds{1}_{l=r, k=s} \Big) \prod_{j \in \partial k \setminus \{i, l\}} \Big(1 - \alpha_{jk} \cdot p^c_{j \rightarrow k}(t-1) \Big),
\end{equation}
which comes from differentiating messages (\ref{eq:message_a}).
Finally, we are able to compute the derivative in (\ref{eq:derivative}) using:
\begin{equation}
    \begin{split}
        \frac{\partial \mu^c_k(t)}{\partial \alpha_{rs}}
        &= \big( 1 - \bar{p}^c_k \big) \sum_{l \in \partial k} \Big( \alpha_{lk} \cdot \phi^{rs}_{l \rightarrow k}(t-1) + p^c_{l \rightarrow k}(t-1) \mathds{1}_{l=r, k=s} \Big) \prod_{j \in \partial k \setminus l} \Big(1 - \alpha_{jk} \cdot p^c_{j \rightarrow k}(t-1) \Big) \mathds{1}_{(t < T)} \\
        &- \big( 1 - \bar{p}^c_k \big) \sum_{l \in \partial k} \Big( \alpha_{lk} \cdot \phi^{rs}_{l \rightarrow k}(t-2) + p^c_{l \rightarrow k}(t-2) \mathds{1}_{l=r, k=s} \Big) \prod_{j \in \partial k \setminus l} \Big(1 - \alpha_{jk} \cdot p^c_{j \rightarrow k}(t-2) \Big) \mathds{1}_{(t > 0)}.
    \end{split}
\end{equation}
The derivative (\ref{eq:derivative}) can then be used instead of $\frac{\partial \mathcal{L}}{\partial \alpha_{ij}}$ equation (\ref{eq:step_a}) and iterated until convergence in the same manner as we did in the SLICER's framework.

\section{Maximum Likelihood approach for the full observation case}
\label{appendix:4}

For completeness, we explain the maximum likelihood implementation that we used as a baseline in the Main Text for comparisons in the case of full observability. The probability of observed data $\Sigma = \bigcup_{c = 1}^{M}\Sigma^c$, where $\Sigma^c \equiv \{ \tau^c_i \}_{i \in V}$, given a set of spreading parameters $\Omega \equiv \{ \alpha_{ij} \}_{(i,j) \in E}$, can be computed using the following form factorised over individual nodes:
\begin{equation}
    P(\Sigma | \Omega) = \prod_{c \in C} \prod_{i \in V} P_i(\tau^c_i | \Sigma^c, \Omega),
    \label{eq:likelihood}
\end{equation}
with
\begin{equation}
    P_i(\tau_i^c | \Sigma^c, \Omega) = \Bigg(\prod_{k \in \partial i} (1 - \alpha_{ki} \mathbf{1}_{\tau_k^c \leq \tau_i^c - 2}) \Bigg) \cdot \Bigg( 1 - \prod_{k \in \partial i} (1 - \alpha_{ki} \mathbf{1}_{\tau_k^c = \tau_i^c - 1}) \mathbf{1}_{\tau_i^c < T} \Bigg),
    \label{eq:local_likelihood}
\end{equation}
where $\partial i$ denotes the set of neighbors of node $i$. Note that equation (\ref{eq:local_likelihood}) is only valid for $\tau_i^c > 0$, otherwise the probability is equal to $1$. This allows to get an estimate of the spreading parameters $\Omega^* \equiv \{ \alpha_{ij}^* \}_{(i,j) \in E}$ by solving a convex optimisation problem
\begin{equation}
    \Omega^* = \arg \min_{\Omega} \left( -\log P(\Sigma | \Omega) \right),
    \label{eq:ML1}
\end{equation}
under constraints:
\begin{equation}
    0 \leq \alpha_{ij} \leq 1 \quad \forall_{(i, j) \in E}.
    \label{eq:ML2}
\end{equation}
The results presented in the Main Text were obtained with the Ipopt solver \cite{wachter2006implementation}, using the implementation of \eqref{eq:ML1}-\eqref{eq:ML2} within the JuMP framework for non-linear optimization \cite{dunning2017jump}.

\section{Details of the multi-replica approach}
\label{appendix:5}

\subsection{Learning framework for the replicas}

Here, we provide a detailed algorithm that we used for constructing multi-replica effective mpdels in the Main Text. In order to account for the error in the DMP equations that result from the presence of loops, instead of having one network with a set of parameters $\{ \alpha_{ij} \}$ we consider a set $R$ of structurally identical networks with a unique set of parameters $\{ \alpha^r_{ij} \}$ each. In such setting marginal probability $\mu^s_i(t)$ of activating node $i$ at time $t$ under cascade starting from node $s$ is equal to an average over marginals in each replica. This leads to a following objective function presented in the Main Text and re-stated here for consistency:
\begin{equation}
    \mathcal{O}^{\text{mixture}} = \sum_{s \in S} \sum_{i \in O} \sum_{\tau_i^s} m^{\tau^s_i} \log \left( \frac{1}{|R|} \sum_{r \in R} \mu^s_{i, r}(\tau^s_i) \right).
    \label{eq:replicas_a}
\end{equation}
This objective can be rewritten using approximate marginals obtained separately from DMP equations for each replica:
\begin{equation}
    \mathcal{O}^{\text{mixture}} = \sum_{s \in S} \sum_{i \in O} \sum_{\tau^s_i} m^{\tau^s_i} \log\Bigg( \frac{1}{|R|} \sum_{r \in R} \Big(p^{r,s}_i(\tau^s_i) \cdot \mathds{1}_{(\tau^s_i < T)} - p^{r,s}_i(\tau^s_i - 1) \cdot \mathds{1}_{(\tau^s_i > 0)} + \mathds{1}_{(\tau^s_i = T)}\Big) \Bigg),
\end{equation}
where $p^{r,s}_i(t)$ is the marginal probability of node $i$ being activated at time $t$ or earlier by a cascade originated from $s$, under replica $r$. Following the procedure described for SLICER in \ref{sec:framework} we compute a Lagrangian for the multi-replica objective function. We will require separate sets of Lagrange multipliers for each replica. We define $\lambda^{r,s}_i(t)$ and $\lambda^{r,s}_{i \rightarrow j}(t)$ as the Lagrange multipliers for marginal probability $p^{r,s}_i(t)$ and message $p^{r,s}_{i \rightarrow j}(t)$ equations respectively. The update equations for the multipliers can be computed using the derivative of the Lagrangian $\mathcal{L}$ over the DMP marginals:
\begin{equation}
\begin{split}
    \frac{\partial \mathcal{L}}{\partial p^{r,s}_i(t)} &= \lambda^{r,s}_i(t) + \sum_{\tau_i^s} \frac{m^{\tau^s_i} \cdot \mathds{1}_{(t=\tau^s_i)} \cdot \mathds{1}_{(\tau^s_i < T)}}{\sum_{r \in R} \Big(p^{r,s}_i(\tau^s_i) - p^{r,s}_i(\tau^s_i - 1) \cdot \mathds{1}_{(\tau^s_i > 0)}\Big)} \\
    &+ \sum_{\tau_i^s} \frac{m^{\tau^s_i} \cdot \mathds{1}_{(t=\tau^s_i-1)} \cdot \mathds{1}_{(\tau^s_i > 0)}}{\sum_{r \in R} \Big(p^{r,s}_i(\tau^s_i - 1) - p^{r,s}_i(\tau^s_i) \cdot \mathds{1}_{(\tau^s_i < T)} - \mathds{1}_{(\tau^s_i = T)}\Big)}.
\end{split}
\end{equation}
The message-multipliers can be obtained using the derivatives over DMP messages and previously obtained marginal-multipliers:
\begin{equation}
    \begin{split}
        \frac{\partial \mathcal{L}}{\partial p^{r,s}_{i \rightarrow j}(t)} &= \lambda^{r,s}_{i \rightarrow j}(t) - \lambda^{r,s}_j(t+1) \, \alpha^r_{ij} \, \big( 1 - \bar{p}^s_j \big) \prod_{m \in \partial j \setminus i} \Big(1 - \alpha^r_{mj} \cdot p^{r,s}_{m \rightarrow j}(t) \Big) \\
        &- \sum_{k \in \partial j \setminus i} \lambda^{r,s}_{j \rightarrow k}(t+1) \, \alpha^r_{ij} \, \big( 1 - \bar{p}^s_j \big) \prod_{m \in \partial j \setminus \{ i, k \}} \Big(1 - \alpha^r_{mj} \cdot p^{r,s}_{m \rightarrow j}(t) \Big).
    \end{split}
\end{equation}
Finally one can use the Lagrange multipliers to compute the derivatives over the spreading parameters:
\begin{equation}
    \begin{split}
        \frac{\partial \mathcal{L}}{\partial \alpha^r_{ij}} =
        &-\sum_{s \in S} \sum_{t=0}^{T-1} \lambda^{r,s}_j(t+1) \, p^{r,s}_{i \rightarrow j}(t) \, \big( 1 - \bar{p}^s_j \big) \prod_{m \in \partial j \setminus i} \Big(1 - \alpha^r_{mj} \cdot p^{r,s}_{m \rightarrow j}(t) \Big) \\
        &-\sum_{s \in S} \sum_{t=0}^{T-1} \sum_{k \in \partial j \setminus i} \lambda^{r,s}_{j \rightarrow k}(t+1) \, p^{r,s}_{i \rightarrow j}(t) \, \big( 1 - \bar{p}^s_j \big) \prod_{m \in \partial j \setminus \{ i, k \}} \Big(1 - \alpha^r_{mj} \cdot p^{r,s}_{m \rightarrow j}(t) \Big).
    \end{split}
    \label{eq:alpha_r}
\end{equation}
In case of non-zero parameters $\alpha^r_{ij}$, the above equation can be significantly simplified:
\begin{equation}
    \frac{\partial \mathcal{L}}{\partial \alpha^r_{ij}} = - \frac{1}{\alpha^r_{ij}} \sum_{s \in S} \sum_{t=0}^{T-1} \left( \lambda^{r,s}_{i \rightarrow j}(t) \, p^{r,s}_{i \rightarrow j}(t) + \lambda^{r,s}_{j \rightarrow i}(t) \, p^{r,s}_{j \rightarrow i}(t) \right).
\end{equation}
As before, the above derivatives serve as a gradient to update parameters $\alpha^r_{ij}$:
\begin{equation}
    \alpha^r_{ij} \longleftarrow \alpha^r_{ij} + \varepsilon \cdot \frac{\partial \mathcal{L}}{\partial \alpha^r_{ij}}.
\end{equation}
Notice that for a single replica, this procedure reduces to the SLICER algorithm described in section \ref{sec:framework}.

\subsection{Remarks on the iterative improvement of the effective model}

We present empirical tests showing that due to the non-convexity of the problem, random initializations lead to different local solutions with a similar prediction accuracy. This phenomenon is illustrated in Fig.~\ref{fig:init}, where we test $10$ different initial conditions (here, for a single-replica case).     

\begin{figure}[ht]
    \centering
    \includegraphics[width=0.75\columnwidth]{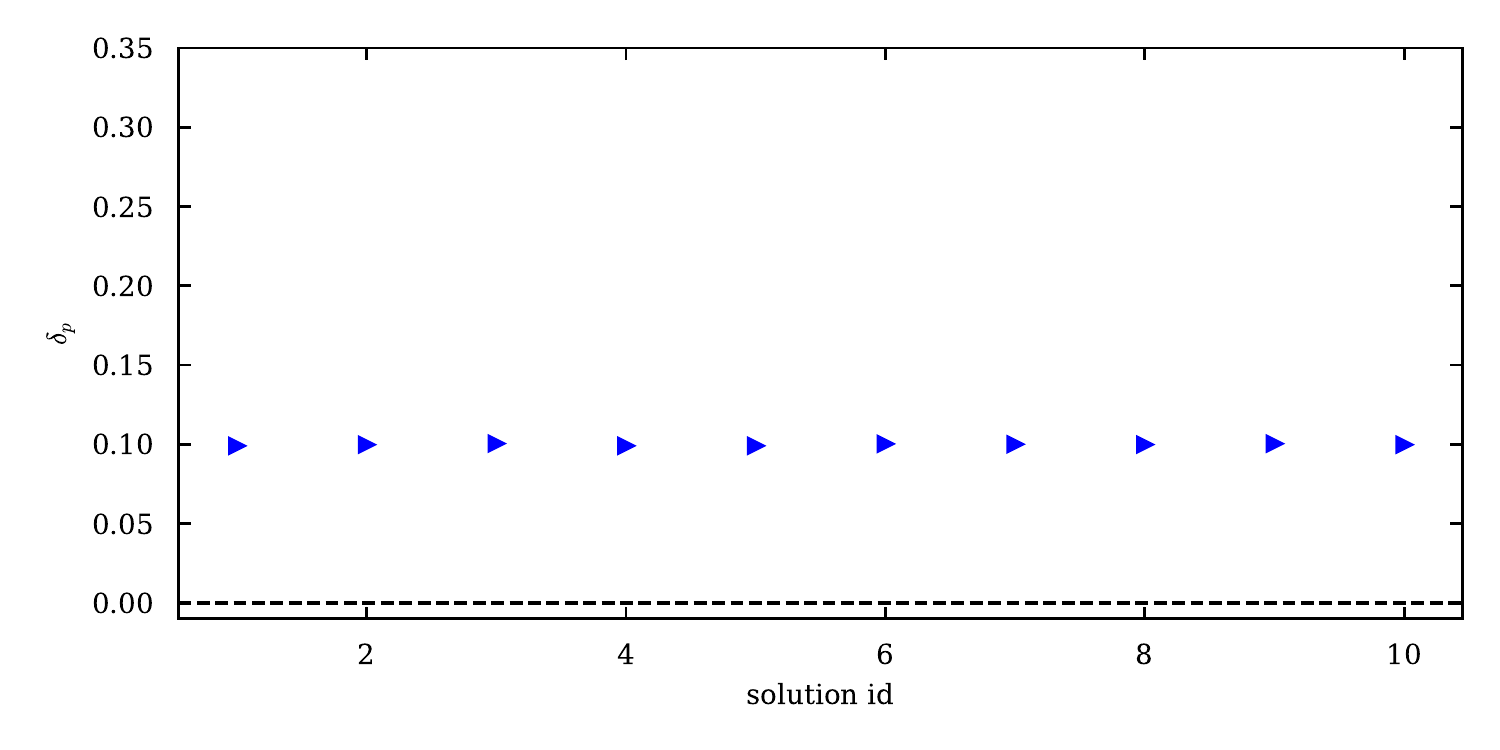}
    \caption{Marginal prediction error for different random initialisation of the single-replica effective model. The results were obtained for a square lattice with $N = 100$ nodes and $M = 10^6$ cascades of length $T = 20$.}
    \label{fig:init_a}
\end{figure}

\begin{figure}[ht]
    \centering
    \includegraphics{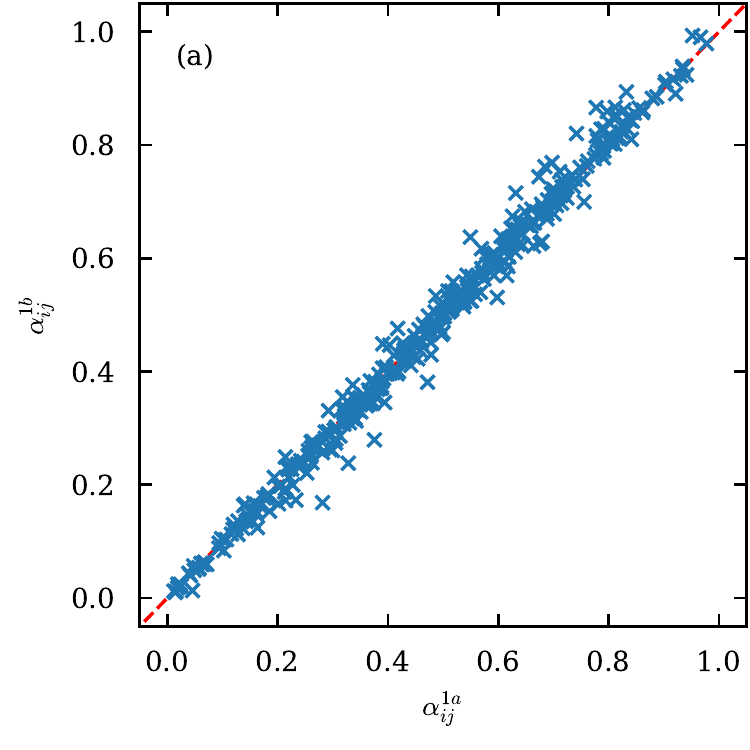}
    \includegraphics{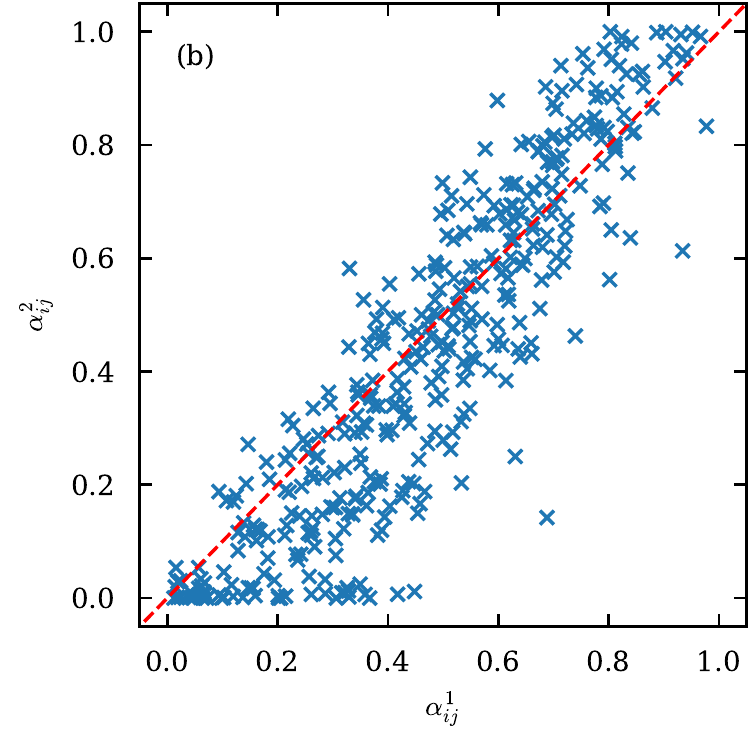}
    \caption{Scatter plots of parameters $\alpha_{ij}$ obtained for different replicas. The results were obtained for a square lattice with $N = 100$ nodes and $M = 10^6$ cascades of length $T = 20$. (a) Comparison of parameters inferred for a single-replica model with different SLICER initializations. (b) Comparison of one of the layers of the 2-replica solution and the single-replica model used as an initialization to the 2-replica SLICER algorithm.}
    \label{fig:scatter}
\end{figure}

A scatter plot of two solutions obtained with different random initializations is shown in Fig. \ref{fig:scatter}(a), where we see that although the reconstructed parameters are different, they are strongly aligned with each other. This observation motivates the multi-replica approach, which suggests that instead of spending the computational budget by warm-starting SLICER multiple times, it is beneficial to invest computing time into improving a given single-replica solution. Fig. 6 of the Main Text suggested that for a fixed number of replicas, an iterative improvement of the solution starting with an initialization given by a slightly perturbed (to break the symmetry) solution obtained with for an effective model with a smaller number of replicas is more advantageous compared to the search of an optimal multi-replica solution using random initialization. Indeed, this procedure guarantees the improvement of the solution as long as the multi-replica objective is maximized, while the search of the parameters of the multi-replica model from random initialization typically leads to a lower-quality solution as the gradient descent is likely stuck at a local optimum (this was the case in all test experiments we ran). Fig. \ref{fig:scatter}(b) shows an example of two replica solution (one of its layers) against a single replica solution that was used as initialization. We see that the parameters diverge, trying to compensate errors arising from the loops in the graph and better match the empirical marginal probability distributions.

\end{document}